\documentclass[11pt,twoside,nohyper]{pallis}

\usepackage{epsfig}
\usepackage{amssymb}
\usepackage{latexsym}
\usepackage{times}
\usepackage{slashed,cite}
\usepackage{upgreek}
\usepackage{rotating}
\usepackage{amsmath}
\usepackage{amsfonts}
\usepackage{amsbsy}
\usepackage{amscd}
\usepackage{bbm}
\usepackage{fancyh}
\usepackage{graphics}
\usepackage{multirow}

\newcommand{\bal}{\begin{align}}
\newcommand{\eal}{\end{align}}
\newcommand{\beqs}{\begin{subequations}}
\newcommand{\eeqs}{\end{subequations}}
\newcommand{\eec}{\end{center}}
\newcommand{\bec}{\begin{center}}

\newcommand{\ecs}{\end{cases}}
\newcommand{\bcs}{\begin{cases}}

\newcommand{\eem}{\end{matrix}}
\newcommand{\bem}{\begin{matrix}}
\newcommand{\eeq}{\end{equation}}
\newcommand{\beq}{\begin{equation}}
\newcommand{\ba}{\begin{array}}
\newcommand{\ea}{\end{array}}
\newcommand{\bea}{\begin{eqnarray}}
\newcommand{\eea}{\end{eqnarray}}
\newcommand{\baq}{\begin{eqnarray}}
\newcommand{\eaq}{\end{eqnarray}}

\newcommand\eqs[2]{Eqs.~(\ref{#1}) and (\ref{#2})}

\newcommand\eqsss[4]{Eqs.~(\ref{#1}), (\ref{#2}), (\ref{#3}) and (\ref{#4})}

\newcommand{\ftn}{\footnotesize}

\newcommand{\TeV}{{\mbox{\rm TeV}}}
\newcommand{\MeV}{{\mbox{\rm MeV}}}
\newcommand{\GeV}{{\mbox{\rm GeV}}}

\newcommand{\eV}{{\mbox{\rm eV}}}
\newcommand{\ZeV}{{\mbox{\rm ZeV}}}
\newcommand{\EeV}{{\mbox{\rm EeV}}}
\newcommand{\PeV}{{\mbox{\rm PeV}}}
\newcommand{\YeV}{{\mbox{\rm YeV}}}

\newcommand{\sFref}[2]{Fig.~\ref{#1}-{\ftn\sf ({#2})}}

\newcommand{\fref}[1]{Fig.~\ref{#1}}

\newcommand{\Eref}[1]{Eq.~(\ref{#1})}
\newcommand{\Sref}[1]{Sec.~\ref{#1}}
\newcommand{\Fref}[1]{Fig.~\ref{#1}}
\newcommand{\Tref}[1]{Table~\ref{#1}}
\newcommand{\cref}[1]{Ref.~\cite{#1}}

\newcommand{\etal}{{\it et al.\/}}

\def\to{\rightarrow}

\def\lf{\left(}
\def\rg{\right)}
\newcommand\vev[1]{\langle {#1} \rangle}
\newcommand\vevi[1]{\langle {#1} \rangle_{\rm I}}
\newcommand\vevii[1]{\left\langle { #1} \right\rangle_{\rm I}}
\newcommand{\Gr}{\ensuremath{\widetilde{G}}}
\newcommand{\Yb}{\ensuremath{Y_{B}}}
\newcommand{\Yg}{\ensuremath{Y_{3/2}}}
\newcommand{\Vhi}{\ensuremath{\widehat{V}_{\rm IPI}}}

\newcommand{\Vm}{\ensuremath{\widehat{V}_{\rm m}}}

\newcommand{\Hhi}{\ensuremath{\widehat{H}_{\rm IPI}}}
\newcommand{\what}{\ensuremath{\widehat}}
\newcommand{\wtilde}{\ensuremath{\widetilde}}
\newcommand{\Khi}{\ensuremath{K}}
\newcommand{\Whi}{\ensuremath{W}}
\newcommand{\Vhio}{\ensuremath{\widehat{V}_{\rm IPI0}}}
\newcommand{\Ns}{\ensuremath{\widehat{N}_\star}}

\newcommand{\mP}{\ensuremath{m_{\rm P}}}
\newcommand{\Mgut}{\ensuremath{M_{\rm GUT}}}
\newcommand{\mgut}{\ensuremath{M_\mathbb{G}}}
\newcommand{\Qef}{\ensuremath{\Lambda_{\rm UV}}}
\newcommand{\Ggut}{\ensuremath{\mathbb{G}}}

\newcommand{\Gbl}{\ensuremath{\mathbb{G}_{B-L}}}
\newcommand{\Gsm}{\ensuremath{\mathbb{G}_{\rm SM}}}

\newcommand{\lm}{\ensuremath{\lambda_\mu}}

\def\openone{\leavevmode\hbox{\small1\kern-3.8pt\normalsize1}}
\newcommand{\dV}{\ensuremath{\Delta V_{\rm I}}}

\newcommand{\fw}{\ensuremath{f_{\rm W}}}

\newcommand{\nst}{\ensuremath{N_{\rm st}}}
\newcommand{\nb}{\ensuremath{N}}

\newcommand{\ca}{\ensuremath{c_{\cal R}}}
\newcommand{\ck}{\ensuremath{c_{\cal R}}}

\newcommand{\fr}{\ensuremath{f_{\cal R}}}
\newcommand{\fk}{\ensuremath{f_{\rm K}}}

\newcommand{\hr}{\ensuremath{F_{\cal R}}}
\newcommand{\hsh}{\ensuremath{F_{\rm sh}}}
\newcommand{\kb}{\ensuremath{K_{\rm st}}}
\newcommand{\ka}{\ensuremath{K_{\rm I}}}

\newcommand{\Omg}{\ensuremath{\Omega}}

\newcommand{\Gsn}{\ensuremath{{\Gamma}_{\rm \dph}}}
\newcommand{\GNsn}{\ensuremath{{\Gamma}_{\dph\to N_i^c}}}
\newcommand{\Ghsn}{\ensuremath{{\Gamma}_{\dph\to H}}}
\newcommand{\Gysn}{\ensuremath{\Gamma_{\dph\to XYZ}}}
\newcommand{\msn}{\ensuremath{m_{\rm \dph}}}
\newcommand{\mx}{\ensuremath{m_{\tilde\chi}}}

\newcommand{\nta}{\ensuremath{\tilde\chi}}
\newcommand{\Omx}{\ensuremath{\Omega_{\tilde\chi}h^2}}
\newcommand{\mss}{\ensuremath{\widetilde m}}
\newcommand{\aS}{\ensuremath{{\rm a}_S}}
\newcommand{\Ald}{\ensuremath{A_\lambda}}

\newcommand{\hd}{{\ensuremath{H_d}}}
\newcommand{\hu}{{\ensuremath{H_u}}}
\newcommand{\ssni}{\ensuremath{\widetilde N^c_i}}
\newcommand{\sni}{\ensuremath{N^c_i}}

\newcommand{\ks}{\ensuremath{k_\star}}
\newcommand{\ns}{\ensuremath{n_{\rm s}}}
\newcommand{\as}{\ensuremath{a_{\rm s}}}
\newcommand{\As}{\ensuremath{A_{\rm s}}}

\newcommand{\rcc}{\ensuremath{\mathcal{R}}}
\newcommand{\rce}{\ensuremath{\widehat{\mathcal{R}}}}
\newcommand{\Ve}{\ensuremath{\widehat V}}

\newcommand{\He}{\ensuremath{\widehat {H}}}

\newcommand{\geu}{\ensuremath{\widehat g}}
\newcommand{\eph}{\ensuremath{\widehat \epsilon}}
\newcommand{\ith}{\ensuremath{\widehat \eta}}

\newcommand{\dph}{\ensuremath{\delta\phi}}

\def\ve{\varepsilon}
\def\aal{{\bar\alpha}}
\def\bbet{{\bar\beta}}
\def\al{{\alpha}}

\def\th{{\theta}}
\def\thb{{\bar\theta}}

\def\thn{{\theta_{\Phi}}}

\newcommand{\Trh}{\ensuremath{T_{\rm rh}}}
\newcommand{\Tns}{\ensuremath{T_{\rm BBN}}}
\newcommand{\Tgr}{\ensuremath{T_{3/2}}}
\newcommand{\Ggr}{\ensuremath{\Gamma_{3/2}}}
\newcommand{\Tf}{\ensuremath{T_{\rm fo}}}
\newcommand{\xf}{\ensuremath{x_{\rm fo}}}

\newcommand{\sg}{\ensuremath{\phi}}

\newcommand{\sgm}{\ensuremath{\phi_{\rm m}}}

\newcommand{\sgmr}{\ensuremath{\phi_{\rm mr}}}
\newcommand{\sgx}{\ensuremath{\phi_\star}}
\newcommand{\sgf}{\ensuremath{\phi_{\rm f}}}
\newcommand{\sgof}{\ensuremath{\phi_{\rm of}}}

\newcommand{\ld}{\ensuremath{\lambda}}
\newcommand{\ldu}{\ensuremath{\uplambda}}
\newcommand{\Ld}{\ensuremath{\Lambda}}

\newcommand{\mgr}{\ensuremath{m_{3/2}}}
\newcommand{\mmgr}{\ensuremath{\mu/m_{3/2}}}

\newcommand{\dW}{\ensuremath{\Delta W}}

\newcommand{\vtau}{\ensuremath{\uptau}}
\newcommand{\dn}{\ensuremath{\delta N}}

\newcommand{\phc}{\ensuremath{\Phi}}
\newcommand{\phcb}{\ensuremath{\bar\Phi}}
\newcommand{\wrh}{\ensuremath{w_{\rm rh}}}
\newcommand{\woi}{\ensuremath{w_{\rm oi}}}

\newcommand{\am}{\ensuremath{{\rm a}_{\mu}}}

\newcommand{\rhna}{\ensuremath{N^c_1}}
\newcommand{\rhni}{\ensuremath{N^c_i}}
\newcommand{\mrh[1]}{\ensuremath{M_{#1N^c}}}
\newcommand{\lrh[1]}{\ensuremath{\lambda_{#1N^c}}}
\newcommand{\mD[1]}{\ensuremath{m_{#1\rm D}}}
\newcommand{\mn[1]}{\ensuremath{m_{#1\rm \nu}}}

\def\Ka{K\"{a}hler potential}
\def\Km{K\"{a}hler manifold}
\def\Ke{K\"{a}hler metric}
\def\Kaa{K\"{a}hler~}
\def\sub{subplanckian}

\newcommand{\plk}{{\it Planck}}

\def\fhi{{IPI}}

\def\actc{{\sf\small P-ACT-LB-BK18}}

\def\actcf{{\sf\ftn P-ACT-LB-BK18}}

\newcommand{\diag}{\ensuremath{{\sf diag}}}

\newcommand{\re}{\ensuremath{{\sf Re}}}

\renewcommand{\arg}{\ensuremath{{\small\sf arg}}}

\title{\LARGE\boldmath \bfseries\scshape Induced-Gravity Palatini-Like Higgs Inflation
in Supergravity Confronts ACT DR6}

\author{\Large \bfseries\scshape C. Pallis\\
School of Technology,\\ Aristotle University of Thessaloniki, \\
Thessaloniki, GR-541 24 GREECE; \\ \vspace{3pt}
\email{kpallis@auth.gr} }

\abstract{We formulate within Supergravity a model of
induced-gravity inflation, excellently consistent with ACT DR6,
inspired by the Palatini gravity. The inflaton belongs in the
decomposition of a conjugate pair of Higgs superfields which lead
to the spontaneous breaking of a $U(1)_{B-L}$ symmetry at a scale
close to the range $(0.145-8.35)\cdot10^{16}~\GeV$. The inflaton
field is canonically normalized thanks to one real and
shift-symmetric contribution into the \Ka. It also includes two
separate holomorphic and antiholomorphic logarithmic terms, the
argument of which can be interpreted as the coupling of the
inflaton to the Ricci scalar. The attainment of inflation allows
for subplanckian inflaton values and energy scales below the
cut-off scale of the corresponding effective theory. Embedding the
model in a ${B-L}$ extension of the MSSM we show how the $\mu$
parameter can be generated and non-thermal leptogenesis can be
successfully realized. An outcome of our scheme is split SUSY with
gravitino mass in the range $(40-60)~\PeV$, which is consistent
with the results of LHC on the Higgs boson mass.

\\ \\ {\ftn\sffamily {\scshape Keywords}:  Cosmology, Inflation, Supersymmetric Models} \\
{\ftn\sffamily {\scshape PACS codes}:  98.80.Cq, 12.60.Jv,
95.30.Cq, 95.30.Sf}\\\\ {\sl\bfseries Published in} {\sl
Astronomy} {\bf 5}, no.~2, 9 (2026)}

\begin{document}

\pagestyle{fancyplain}

\maketitle

\rhead[\fancyplain{}{ \bf \thepage}]{\fancyplain{}{\sl IG
Palatini-like Higgs Inflation in SUGRA Confronts ACT DR6}}
\lhead[\fancyplain{}{\sl C. Pallis}]{\fancyplain{}{\bf \thepage}}
\cfoot{}


\section{Introduction} \label{intro}

The \emph{Data Release 6} ({\sf\ftn DR6}) from the \emph{Atacama
Cosmology Telescope} ({\sf\ftn ACT}) \cite{act,actin} favors a
slightly higher scalar spectral index, $\ns$, than the one
indicated from the \plk\ data \cite{plin} prompting renewed
interest in inflationary models that can accommodate such a shift
\cite{act0,act2,act4,gup,oxf,indi,kina,actlee,rhc,rhb,
rha,act5,nmact,maity,reh8, actattr,act1,actj,yin,actpal,
act3,act6,act7,act8,ketov,
r2a,r2b,r2drees,r2mans,r2li,heavy1,heavy,fhi1,fhi2,fhi3,fhi4,fhi5,smth,okada,ellis10,
ellis8, king, yermek, channuie, sami,zeld,turk,ketov1,waqas,
actpole,phi,r3,laura,heur} -- for a review see \cref{actreview}.
Indeed, the combination of the aforementioned data with the
measurements from other experiments \cite{plin, bcp, desi} too,
named \actc\ data, suggests that $\ns$, its running \as\ and the
tensor-to-scalar ratio $r$, are to be confined at 95$\%$
\emph{confidence level} ({\sf\small c.l.}) in the ranges
\cite{actin}
\beq \label{data} \ns=0.974\pm0.0068,~\as  =
0.0062\pm0.0104~~\mbox{and}~~r\leq0.038. \eeq

In a recent paper \cite{actpal} we propose a variant of
non-minimal inflation \cite{nmi,roest}, characterized as
kinetically modified, adopting the Palatini formulation
\cite{demir, palreview} of gravity. In particular, we employ a
chaotic potential of the form $\sg^n$ for the inflaton field
$\sg$, with $n=2,4$, and a kinetic mixing $\fk=\fr^m$, besides the
non-minimal coupling of the inflaton to gravity
$\fr=1+\ck\sg^{n/2}$. Our setting causes an elevation of $\ns$,
compared to its value in the metric formulation, without an
accompanied $r$ increase, obtained in the last case \cite{nmikr,
expkr,varkr,jhep,nmhkr}. As a consequence, comfortable
compatibility with \Eref{data} can be achieved. An outstanding
feature of our proposal is the possibility to obtain canonically
normalized $\sg$ for $m=1$, which is not easily feasible in the
metric formulation. {Although Palatini gravity has \cite{renap} no
\emph{Supergravity} ({\sf\ftn SUGRA}) embedding, our models in
\cref{actpal} can be realized within standard (Poincar\'e) SUGRA
thanks to the selected \Ka\ with the $m=1$ case assuring low
dimensionality.}



In our present investigation we adapt our proposal in
\cref{actpal} to an elementary \emph{Supersymmetric} ({\sf \ftn
SUSY}) \emph{Grand Unified Theory} ({\sf \ftn GUT}) performing the
following arrangements:

\begin{itemize}

\item[{\sf\ftn (i)}] We promote the inflaton to the radial part of
a conjugate pair of superheavy Higgs superfields. This option
allows us to focus on the quartic inflationary potential taking
$n=4$ in the formulae of \cref{actpal} -- for other works on
GUT-scale Higgs inflation see, e.g., \cref{hi1, hi2, hi3, hi4,
hi5, nmBL, jhep, ighi, ighic, phi, sor, tmhi,
nmhkr,varkr,higgsflaton}.

\item[{\sf\ftn (ii)}] We restrict ourselves to a \Ka\ with just
quadratic contributions. Similarly to the case with gauge-singlet
inflation \cite{actpal}, this second assumption leads to canonical
kinetic terms for the inflaton. As a consequence, we reveal a
model recently analyzed in \cref{un5} with focus on its
consistency with unitarity
\cite{un1,un2,riotto,udemir,un0,un4,un3}.

\item[{\sf\ftn (iii)}] We incorporate the idea of \emph{Induced
Gravity} ({\sf\ftn IG}) \cite{old}  according to which the
(reduced) Planck mass $\mP$ is generated  via the \emph{vacuum
expectation value} ({\ftn\sf v.e.v}) that a scalar field acquires
at the end of a phase transition in the early universe. In our
scheme, the scalar field is identified with the higgsflaton -- cf.
\cref{igi0, igi1, sm, igi2, lee, higgsflaton, R2r, rena, rena1,
igi, igic, igir, ighi,ighic}.

\item[{\sf\ftn (iv)}] We connect the strength of the non-minimal
coupling to gravity with the scale of the unification of the gauge
coupling constants of the \emph{Minimal SUSY Standard Model}
({\sf\small MSSM}) -- cf.~\cref{ighi, ighic}.

\end{itemize}

The hypotheses above give rise to an economical, predictive and
well-motivated set-up, which can be characterized, following the
terminology of \cref{actpal}, as (kinetically modified with $m=1$)
\emph{induced-gravity Palatini-like Higgs inflation} ({\sf\small
IPI}). This construction actually ``ACT-ivates'' our older models
in \cref{ighi, ighic}, which may become consistent with
\Eref{data} only at the cost of some tuning regarding the
coefficients of the logarithm in the \Ka s. This tuning can be
totally avoided in our present model, although we discuss, just
for completeness, the behavior of our results with variation of
the relevant coefficient ($N$).

Apart from the compatibility with \Eref{data}, \fhi\ offers  also
the opportunity to explore consequences beyond the inflationary
phenomenology. Namely, embedding it into a $B-L$ extension of MSSM
we can accommodate a resolution of the $\mu$ problem, following
the mechanism in \cref{ighi,ighic}, and acceptable baryogenesis
via \emph{non-thermal leptogenesis} ({\sf\small nTL})
\cite{dreeslept,zhang,lept} consistently with  gravitino ($\Gr$)
cosmology \cite{brand, kohri,grspanos}. Contrary to previous
implementations of the same post-inflationary scheme -- cf.
\cref{tmhi, phi, ighi, ighic} --, the outcome here includes split
SUSY \cite{split, split1, strumia} with very short-lived $\Gr$.

The remaining text is structured as follows. In Sec.~\ref{ighi} we
present the cornerstones of our proposal and, in \Sref{fhi}, we
analyze the inflationary scenario. We then -- in \Sref{pfhi} --
examine a possible post-inflationary completion of our setting.
Our conclusions are summarized in \Sref{con}. In Appendix
\ref{app} we address a subtlety related to a very short stage of
\emph{oscillating inflation} ({\ftn\sf OI}) \cite{os1,os2,wands,
rhlin} encountered after the end of \fhi. Throughout the text, the
subscript of type $,z$ denotes derivation \emph{with respect to}
({\ftn\sf w.r.t}) the field $z$ and charge conjugation is denoted
by a star. Unless otherwise stated, we use units where $\mP =
2.43\cdot 10^{18}~\GeV$ is taken to be unity.

\section{Higgs Inflation and Induced Gravity} \label{ighi}

In \Sref{set} we specify the relevant super- and \Ka\ of our
proposal, whereas in \Sref{igsec} we explain the imposition of the
IG condition.

\subsection{Set-up} \label{set}

Our starting point is the superpotential known from the models of
F-term hybrid inflation \cite{fhi1,fhi2,fhi3,fhi4,fhi5}
\beq W_{\rm IPI}=\ld S\lf\bar\Phi\Phi-M^2/4\rg\label{whi} \eeq
which is uniquely determined, at renormalizable level, by a gauge
symmetry $\Ggut$ and a convenient continuous $R$ symmetry. Here,
$\ld$ and $M$ are two constants which can both be taken positive
by field redefinitions; $S$ is a left-handed superfield, singlet
under $\Ggut$; $\bar\Phi$ and $\Phi$ is a pair of left-handed
superfields belonging to non-trivial conjugate representations of
$\Ggut$, and reducing its rank by their v.e.vs. Just for
definiteness we restrict ourselves to $\Gbl=\Gsm\times U(1)_{B-L}$
which consists the simplest GUT beyond the MSSM  -- here  $\Gsm$
is the gauge group of MSSM and $B$ and $L$ denote the baryon and
lepton number. With the specific choice of $\Ggut$ the
$U(1)_{B-L}$ and $R$ charges of the various superfields are
\beq (B-L)(S, \phcb, \Phi)=(0,-2,2)~~~\mbox{and}~~~R(S, \phcb,
\Phi)=(2,0,0).\label{qblr}\eeq
At the SUSY limit we expect that $W$ leads to a $B-L$ phase
transition as we show in \Sref{ig2} below.

The realization of \fhi\ with the inflaton included in the
$\phcb-\phc$ system and $S$ stabilized at the origin requires a
careful choice of $K$ -- as, e.g., in \cref{ighi, sor,jhep,nmBL}.
{ The proposed here $K$ includes two contributions without mixing
between the $\phcb-\Phi$ and $S$ sectors. Namely, it is written as
\beq K=\ka+\kb. \label{ktot}\eeq
The second term in $K$ successfully stabilizes $S$ at the origin
without invoking higher order terms -- cf. \cref{linde1}. In
particular, $\kb$ takes the form
\beq \kb=\nst\ln\lf1+{|S|^2/\nst}\rg~~\mbox{with}~~0<\nst<6
\label{kb}\eeq
which parameterizes \cite{su11} the compact manifold $SU(2)/U(1)$
with curvature $2/\nb$. On the other hand, $\ka$ depends on the
$\bar\Phi-\Phi$ sector and includes the functions
%
\beq \label{hr}
\hr(\Phi)=4\ca\phcb\phc~~\mbox{and}~~\hsh=|\Phi-\phcb^*|^2, \eeq
from which \hr\ is holomorphic and can be employed as the
non-minimal coupling to gravity, whereas $\hsh$ is real, assisting
us to incorporate the canonical kinetic mixing. These functions
extend the corresponding ones in \cref{actpal} to the case of a
conjugate pair of gauge non-singlet superfields $\phcb-\phc$. In
sharp contrast to those functions, however, unity is not included
in $\hr$. As explained in \Sref{igsec}, the identification of
$\hr$ with unity  at the vacuum of the theory essentially
encapsulates the IG hypothesis --
cf.~\cref{R2r,igi,igic,ighi,ighic,igir}. With the ingredients
above, we construct $\ka$ in \Eref{ktot} which is
\beq \ka=-\frac{N}{2}\ln\hr-\frac{N}{2}\ln\hr^*+\hsh
\label{ka}\eeq
and features the structure of $\ka$ in \cref{actpal} -- at
renormalizable level, i.e., $m=1$ -- with the introduction of
variable coefficients of the logarithms. It is worth noticing that
contrary to the embedding of IG Higgs inflation in metric SUGRA --
cf. \cref{ighi,ighic} --, $\hr$ and $\hr^*$ enter two different
terms in \Eref{ka} and so they give zero contribution into the
\Ke, which takes the form
\beq \label{k1ab} M_{\phcb\Phi}=\lf K_{\rm I\al\bbet}\rg=\diag\lf
1,1\rg,\eeq
where we use the standard notation $K_{\rm I\al\bbet}=K_{\rm
I,z^\al z^{*\bbet}}$ with $z^\al=\Phi,\phcb$. Alternative forms of
$K$ in \Eref{ktot} as those discussed in \cref{ighi} can be also
constructed using variants of $\kb$ in \Eref{kb} and changing its
placement in the logarithms of $\ka$. We adopt here the simplest
option which enjoys an exact symmetry regarding $\kb$.}


\subsection{Implementation of IG Condition}\label{igsec}

The scale $M$ and the function $\hr$ involved in \eqs{whi}{ka}
assist us in implementing the idea of IG. To explain how it works,
we introduce our notation in the Einstein and Jordan frames in
\Sref{ig1} and then, in \Sref{ig2}, we derive the SUSY vacuum
which plays a key role in imposing the IG condition.

\subsubsection{From Einstein to Jordan Frame}\label{ig1}

The part of the \emph{Einstein frame} ({\sf\ftn EF}) action within
SUGRA related to the complex scalars $z^\al=\phc,\phcb, S$ --
denoted by the same superfield symbol -- has the form
\cite{linde1}
\beqs\beq\label{Saction1}  {\sf S}=\int d^4x \sqrt{-\what{
\mathfrak{g}}}\lf-\frac{1}{2}\rce +K_{\al\bbet} \geu^{\mu\nu}D_\mu
z^\al D_\nu z^{*\bbet}-\Ve\rg\,, \eeq
where $\widehat{\mathfrak{g}}$ is the determinant of the EF metric
$\geu_{\mu\nu}$, $\rce$ is the EF Ricci scalar curvature, $D_\mu$
is the gauge covariant derivative, and $K^{\al\bbet}$ is the
inverse of the \Ke\ defined as
$K^{\al\bbet}K_{\bbet\gamma}=\delta^\al_{\gamma}$. Also, $\Ve$ is
the EF SUGRA potential which includes F- and D- term contributions
which can be found in terms of $\Whi$ and $K$ in \eqs{whi}{ktot}
via the formula
\beq \Ve=e^{\Khi}\left(K^{\al\bbet}(D_\al W_{\rm IPI})D^*_\bbet
W_{\rm IPI}^*-3{\vert W_{\rm IPI}\vert^2}\right)+\frac{g^2}2{\rm
D}_{BL}^2,\label{Vsugra} \eeq\eeqs
where $g$ is the gauge coupling constant -- assumed unified with
the others of \Gsm\ -- and we define the \Kaa covariant derivative
and the D terms due to $\Gbl$ as
\beq D_\al W_{\rm IPI}=W_{{\rm IPI},z^\al}+K_{,z^\al}W_{\rm
IPI}~~\mbox{and}~~{\rm D}_{BL}=\phc K_\phc-\phcb K_{\phcb}=
\lf{|\phc|^2-|\phcb|^2}\rg.\label{dterm} \eeq
Therefore, a field configuration with vanishing ${\rm D}_{BL}$ is
\beq
\vevi{\Phi}=\vevi{\bar\Phi}=\sg/2~~\mbox{and}~~\vevi{S}=0,\label{inftr0}\eeq
where take into account additionally the stabilization of $S$ and
we introduce the symbol $\vevi{...}$ to denote values during IPI.
Henceforth, we confine ourselves to this path -- assuming
furthermore that $\arg(\phc)=\arg(\phcb)$ -- which is a honest
inflationary trajectory -- as we show in \Sref{fhi1} --,
supporting IPI driven exclusively by $\Ve_{\rm F}$.

The action in \Eref{Saction1} can be brought into the \emph{Jordan
frame} (JF) by performing a conformal transformation \cite{linde1,
jhep}. {As clarified in \cref{renap}, SUGRA admits only metric
formulation and, therefore, if we define the JF metric,
$g_{\mu\nu}$, through the relation
\beqs\beq \label{weyl}
\geu_{\mu\nu}=-({\Omega}/{N})g_{\mu\nu},\eeq
where $\Omega$ is the frame function, we are obliged to adopt the
well-known relations between hatted (EF) and unhatted (JF)
quantities mentioned in \cref{jhep, ighi}. As a consequence, we
end up with the following action in the JF
\beq {\sf S}=\int d^4x
\sqrt{-\mathfrak{g}}\lf\frac{\Omega\mP^2}{2N}
\rcc+\frac{3\mP^2}{4N\Omega}D_\mu\Omega D^\mu\Omega
-\frac{1}{N}\Omega K_{\al{\bbet}}D_\mu z^\al D^\mu z^{*\bbet}-V
\rg\>\>\>\mbox{with}\>\>\>V=\frac{\Omg^2}{N^2}\Ve\,.\label{action2}\eeq\eeqs
If, in addition, we connect $\Omega$ to $K$ in Eq.~(\ref{ktot})
through the following relation
\beqs\beq-{\Omega}/N =e^{-K/N\mP^2
}=\hr^{1/2}\hr^{*1/2}e^{-\hsh/N}(1+|S|^2/\nst\mP^2)^{-\nst/N}\label{Omg1}\eeq
we arrive at the following action
\beq {\sf S}=\int d^4x \sqrt{-\mathfrak{g}}\lf\frac{
\Omega}{2N}\mP^2\rcc+\mP^2\lf\Omega_{\al{\bbet}}+\frac{3-N}{N}
\frac{\Omega_{\al}\Omega_{\bbet}}{\Omega}\rg D_\mu z^\al D^\mu
z^{*\bbet}- \frac{27}{N^3}\Omega{\cal A}_\mu{\cal A}^\mu-V \rg,
\label{Sfinal}\eeq
where we take into account the definition \cite{linde1} of the
purely bosonic part of the on-shell value of the auxiliary field
\beq {\cal A}_\mu =i\lf K_\al D_\mu z^\al-K_\aal D_\mu
z^{*\aal}\rg/6=-iN\mP^2 \lf \Omega_\al D_\mu z^\al-\Omega_\aal
D_\mu z^{*\aal}\rg/6\Omega.\label{Acal}\eeq\eeqs
Here we use the shorthand notation $\Omega_\al=\Omega_{,\Phi^\al}$
and $\Omega_\aal=\Omega_{,\Phi^{*\aal}}$.} The first term in the
right-hand side of \Eref{Sfinal} reveals that $-\Omega/N$ plays
the role of a non-minimal coupling to gravity. The emergence of
the conventional Einstein gravity at the vacuum dictates
\beq -\vev{{\Omega}/{N}}=1,\label{igc1}\eeq
where we restore the presence of $\mP$ in the last expressions
above for convenience.


\subsubsection{IG conjecture}\label{ig2}

The IG hypothesis requires the generation of $\mP$ at the vacuum
of the theory, which can be determined expanding $\Ve$ in
\Eref{Saction1} in powers of $1/\mP$. Namely, we obtain the
following low-energy effective potential which plays the role of
the SUSY one
\beqs \beq \label{Vsusy} V_{\rm SUSY}=e^{\widetilde K}\widetilde
K^{\al\bbet} W_{\rm IPI\al} W^*_{\rm IPI\bbet},\eeq
where $\widetilde K$ is the limit of $K$ in Eq.~(\ref{ktot}) for
$\mP\to\infty$.  Namely, we get
\beq \label{Kquad}\widetilde
K=-N\ln(\hr\hr^*)^{1/2}+|\Phi-\phcb^*|^2 +|S|^2,\eeq\eeqs
from which we can then compute the \Ke\ $\widetilde
K_{\al\bbet}=\diag(1,1,1)$. Upon substitution into \Eref{Vsusy} we
obtain
\beq V_{\rm SUSY}\simeq
(\hr\hr^*)^{-N/2}\lf(\ld^2\left|\phcb\phc-{M^2/4}\right|^2+
\ld^2|S|^2\lf|\phc|^2+|\phcb|^2\rg\rg. \label{VF}\eeq
We can easily infer that the SUSY vacuum lies along the direction
\beq \vev{S}=0 \>\>\>\mbox{and}\>\>\>
|\vev{\Phi}|=|\vev{\bar\Phi}|=M/2\>\>\Rightarrow\>\>\vev{\sg}=M,\label{vevs}
\eeq
which is included in the inflationary path of \Eref{inftr1}. As we
see below -- in \Sref{pfhi1} -- $\vev{S}$ may slightly deviate
from its value above after inclusion of soft SUSY-breaking
effects. From \Eref{vevs} it is clear that $\vev{\Phi}$ and
$\vev{\bar\Phi}$ spontaneously break $U(1)_{B-L}$ down to
$\mathbb{Z}^{B-L}_2$. Note that $U(1)_{B-L}$ is already broken
during IPI and so no cosmic strings are formed -- see \Sref{fhi2}.

The scale of $U(1)_{B-L}$ breaking can be related to the magnitude
of $\ck$ via the IG conjecture in \Eref{igc1}. Indeed, upon
substitution of \Eref{vevs} into \Eref{Omg1}, \Eref{igc1} implies
\beq\ck\vev{\sg}^2/\mP^2=1~~\stackrel{(\ref{vevs})}{\Longrightarrow}~~M=\mP/\sqrt{\ck}.
\label{igc2}\eeq
As we see below, both $\ck$ and $M$ can be independently
restricted by the inflationary and the gauge unification
requirements correspondingly enhancing, thereby, the
predictability of the model.

\section{Inflation Analysis}\label{fhi} %

We below -- in \Sref{fhi1} -- derive the inflationary potential of
our model and then present its predictions in \Sref{fhi2}.

\subsection{Inflationary Potential}\label{fhi1}

If we parameterize $\Phi, \bar\Phi$ and $S$ as follows
\beq\label{hpar} \Phi=\frac1{\sqrt{2}}\sg\,
e^{i\th}\cos\thn,~~\bar\Phi=\frac1{\sqrt{2}}
\sg\,e^{i\thb}\sin\thn~~\mbox{where}~~
0\leq\thn\leq\frac{\pi}{2}~~\mbox{and}~~S=\frac1{\sqrt{2}}\lf{s
+i\bar s}\rg, \eeq
the D-flat direction in \Eref{inftr0} can be now expressed as
\beq \label{inftr} \vevii{s}=\vevii{\bar
s}=\vevii{\th}=\vevii{\thb}=0\>\>\mbox{and}\>\>\vevii{\thn}={\pi/4}\,.\eeq
Along this, the only surviving term of $\Ve$ in \Eref{Vsugra} can
be written as
\beqs \beq\label{Vhi} \Vhi= e^{K}K^{SS^*}\, |W_{{\rm IPI},S}|^2=
\frac{\ld^2(\sg^2-M^2)^2}{16\fr^{N}}=\frac{\ld^2\fw^2}{16\ck^{2+N}\sg^{2N}},\eeq
where we define the following quantities
\beq \label{dndef}
\fr=-\vevii{{\Omega}/{N}}=\ck\sg^2,~~\fw=\ck\sg^2-1~~\mbox{and}~~N=2(1+\dn).\eeq\eeqs
{ Note that we make use of \Eref{igc2} for the derivation of the
last expression in \Eref{Vhi}. This one clearly reveals the fact
that the minimum of $\Vhi$ is not placed at zero -- as in the
original Starobinsky model \cite{igir} -- but at $\sg=\vev{\sg}$
given in \Eref{vevs}. This is evidently due to the adoption of the
IG conjecture \cite{R2r, igi, igic} which allows us to obtain a
Starobinsky-like potential. Indeed, for $N=2$ $\Vhi$ develops an
inflationary plateau, with potential energy density $\Vhio$ and
corresponding Hubble parameter $\Hhi$ given respectively by
\beq \Vhio\simeq\frac{\ld^2}{16\ck^2}\,\>\>\mbox{and}\>\>\>
\Hhi\simeq{\Vhio^{1/2}\over\sqrt{3}}={\ld\over4\ck\sqrt{3}},
\label{Vhio}\eeq
in accordance with the well-known Starobinsky-like inflationary
models \cite{R2r, igi, igir}. However, $\dn$ allows for slight
deviations from the absolute flatness as in the cases analyzed in
\cref{igi2,ighi,ighic}. In contrast to those metric realizations
of Starobinsky inflation, in our present case, the inflaton (and
$S$) is canonically normalized since $K_{\al\bbet}$ along the path
in \Eref{inftr} is trivial, i.e.,
\beq \label{kab} \vevi{K_{\al\bbet}}=\diag\lf 1,1,1\rg.\eeq
This is a consequence of the shift-symmetric real part in
\Eref{ka} which can be interpreted as a kinetic modification of
the Palatini inflation motivated in \cref{actpal} and analyzed
recently in \cref{un5} with focus on the issue of unitarity.
Adopting the notation of \cref{actpal} we here use $m=1$ and not
$m=0$ which corresponds to the original (i.e., kinetically
unmodified) model of Palatini inflation \cite{palreview, demir}.
Needless to say, the derivation of the Starobinsky potential is
here based exclusively on the structure of $\Ve$ in \Eref{Vsugra}
and not on a specific relation between the initial and the
canonically normalized inflaton as in case of E- \cite{tkref, emd}
or T-model \cite{eno7,tmd,unvr2}.}


\renewcommand{\arraystretch}{1.2}
 \begin{table}[!t] \bec\begin{tabular}{|c|c|c|c|}\hline
{\sc Fields}&{\sc Eingestates} & \multicolumn{2}{c|}{\sc Masses
Squared}\\ \hline\hline
4 Real &$\theta_{+}$&$m_{\theta+}^2$&{$3(1+4\ck/\fw^2)\Hhi^2$}
\\
Scalars&$\theta_\Phi$ &$m_{
\theta_\Phi}^2$&{$M^2_{BL}+6(1-4\ck/\fw+4(1+\dn)/\sg^2)\Hhi^2$}\\
&$s, \bar{s}$ &$m_{s}^2$&{$6(1/\nst+4/\sg^2\fw^2)\Hhi^2$}\\\hline
1 Gauge Boson& $A_{BL}$ &$ M_{BL}^2$&{$g^2\sg^2$}\\\hline
$7$ Weyl & ${\psi}_\pm$ & $m^2_{\psi\pm}$
&{$24(1-\dn\fw)^2\Hhi^2/\fw^2\sg^2$}\\
Spinors &$\ldu_{BL}, \psi_{\Phi-}$&
$M_{BL}^2$&{$g^2\sg^2$}\\\hline
\end{tabular}\eec
\hfill \caption{\sl\small The mass-squared spectrum of our model
along the configuration in \Eref{inftr}.}\label{tab1}
\end{table}
\renewcommand{\arraystretch}{1.}

We can verify that the inflationary direction in \Eref{inftr} is
stable w.r.t the fluctuations of the fields
$\chi^\al=\theta_+,\theta_\Phi$ and $S$ where
$\th_{\pm}=\lf\bar\th\pm\th\rg/\sqrt{2}$. To this end, we
construct the mass-squared spectrum of those scalars and arrange
the found expressions in \Tref{tab1}. These expressions assist us
to appreciate the role of $\nst>0$ in retaining positive $m^2_S$.
Moreover, $m^2_{\chi^\al}\gg\Hhi^2=\Vhio/3$ for
$\sgf\leq\sg\leq\sgx$ -- the field values $\sgx$ and $\sgf$ are
specified in \Sref{fhi2a} below. In \Tref{tab1} we display also
the masses, $M_{BL}$, of the gauge boson $A_{BL}$ and the masses
of the corresponding fermions with the eigenstate $\what \psi_\pm$
defined as $\psi_\pm =({\psi}_{\Phi+}\pm {\psi}_{S})/\sqrt{2}$.
Since $\Gbl$ is broken during \fhi, no cosmic strings are formed
after its termination.

The derived mass spectrum can be employed in order to find the
one-loop radiative corrections $\dV$ to $\Vhi$. Considering SUGRA
as an effective theory with cutoff scale equal to $\mP$, the
well-known Coleman-Weinberg formula can be employed
self-consistently taking into account only the masses which lie
well below the UV cut-off scale $\Qef$ -- see \Sref{fhi2b} below
-- i.e., all the masses arranged in \Tref{tab1} besides $M_{BL}$
and $m_{\th_\Phi}$ -- cf. \cref{nmBL, jhep}. The resulting $\dV$
takes the form
\beq\label{vrc} \dV=\frac{1}{64\pi^2}\sum_\al {{\rm N}_\al}
m_\al^4\ln \frac{m_\al^2}{\Ld_{\rm CW}^2}~~\mbox{where}~~\bcs
\al&=\{\theta_+,s,\psi_{\Phi\pm}\}\\{\rm
N}_\al&=\{1,2,-4\}\ecs\eeq
and lets intact our inflationary outputs, provided that the
renormalization-group mass scale $\Lambda_{\rm CW}$, is determined
by requiring $\dV(\sgx)=0$ or $\dV(\sgf)=0$. Specific values of
$\Ld_{\rm CW}$ are given in \Sref{num} below.

\subsection{Inflationary Observables -- Constraints}\label{fhi2}

The constraints imposed on our inflationary setting may be grouped
in two categories (observational and theoretical) which are
described respectively in \Sref{fhi2a} and \ref{fhi2b} below.

\subsubsection{Observational Constraints}\label{fhi2a}

These constraints reconcile our model with the present data
\cite{actin}. Namely, they are related to:

\paragraph{(a) Normalization of the Power Spectrum and Number of e-foldings.}
The amplitude $\As$ of the power spectrum of the curvature
perturbations generated by $\sg$ and the number of e-foldings
$\Ns$ that the scale $\ks=0.05/{\rm Mpc}$ experiences during \fhi\
can be computed using the standard formulae  \cite{review}
\begin{equation}
\label{Nhi}  \As^{1/2}= \frac{1}{2\sqrt{3}\, \pi} \;
\frac{\Vhi^{3/2}(\sgx)}{\left|\Ve_{\rm
IPI,\sg}(\sgx)\right|}\>\>\>\mbox{and}\>\>\>\Ns=\int_{\sgf}^{\sgx}
d\sg\frac{\Vhi}{\Ve_{\rm IPI,\sg}},\eeq
where $\sgx$ is the value of $\sg$ when $\ks$ crosses outside the
inflationary horizon, and $\sgf$ is the value of $\sg$ at the end
of IPI, which can be found, in the slow-roll approximation, from
the condition
\beq{\ftn\sf
max}\left\{\eph(\sg),\left|\ith(\sg)\right|\right\}\simeq1,\>\mbox{where}\>\>
\eph=\frac12\left(\frac{\Ve_{\rm IPI,\sg}}{\Ve_{\rm
IPI}}\right)^2\>\>\>\mbox{and}\>\>\> \ith={\Ve_{\rm
IPI,\sg\sg}\over\Ve_{\rm IPI}}\cdot \label{srcon} \eeq
The observables above are to be confronted with the \cite{actin}
\begin{align} \nonumber
&\sqrt{\As}\simeq4.618\cdot10^{-5}\>\>\>\mbox{and}\>\>\>\Ns\simeq61.5+
\ln{\Vhi(\sgx)^{1/2}\over\Vhi(\sgof)^{1/4}}-\frac{1}{3(1+\woi)}\ln
\frac{\Vhi(\sgf)}{\Vhi(\sgof)}\\&+ \frac{1-3\wrh}{6(1+\wrh)}\lf\ln
\lf{\rho_{\rm
rh}}/{\Vhi(\sgof)}\rg^{1/2}-\ln\fr(\sgf)\rg+\frac12\fr(\sgx).
\label{prob}\end{align}
Here, we take into account -- see Appendix \ref{app} -- that IPI
is followed in turn by a very short period of OI and a phase of
damped oscillations with mean equation-of-state parameters
$\woi\simeq-0.71$ and $\wrh\simeq-0.02$ respectively. Note that
$\wrh$ is approximately equal to the value we obtain for a
quadratic potential \cite{turner,dim} and not quartic as naively
expected from the form of numerator of $\Vhi$ in \Eref{Vhi}. Also
$\Trh$ is the temperature after IPI with correspoding energy
density
\beq \rho_{\rm rh}=\pi^2g_{\rm rh*}\Trh^4/30
~~\mbox{where}~~g_{\rm rh*}=228.75 \label{rhor} \eeq
is the energy-density effective number of degrees of freedom for
the MSSM spectrum at the reheat temperature $\Trh$. After
reheating we obtain the well-known radiation and matter
domination.

\paragraph{(b) Remaining Observables.} The compatibility with the
data in \Eref{data} can be checked if we compute the corresponding
inflationary observables through the relations:
\beq\label{ns} \mbox{\ftn\sf (a)}\>\>\> \ns=\:
1-6\widehat\epsilon_\star\ +\
2\widehat\eta_\star,\>\>\>\mbox{\ftn\sf (b)}\>\>\>
\as=\:2\left(4\widehat\eta_\star^2-(\ns-1)^2\right)/3-2\widehat\xi_\star\>\>\>\mbox{and}\>\>\>\mbox{\ftn\sf
(c)}\>\>\>r=16\widehat\epsilon_\star, \eeq
where $\widehat\xi={\Ve_{\rm IPI,\sg} \Ve_{\rm
IPI,\sg\sg\sg}/\Ve_{\rm IPI}^2}$ and the variables with subscript
$\star$ are evaluated at $\sg=\sg_\star$.

\subsubsection{Theoretical Considerations}\label{fhi2b}

We may qualify better our models by taking into account three
additional restrictions of theoretical origin. In particular, we
consider the following issues:

\paragraph{(a) Connection With the Gauge Coupling
Unification}

The $U(1)_{B-L}$ gauge symmetry does not generate any extra
contribution to the renormalization group running of the gauge
coupling constants of $\Gsm$ and so the scale $M$ and the relevant
gauge coupling constant, $g$ can be different than the values
dictated by the MSSM unification of the gauge coupling constants
which gives $g\simeq0.7$ with $\mgut\simeq20~\YeV$. Note that this
result is valid also in case of split SUSY \cite{split1} which
fits well with our scheme -- see \Sref{pfhi1}. In our approach we
mostly keep the $g$ value above and let $\mgut$ varying. The
connection with IG condition can be done if we identify $\mgut$
with the mass of the gauge boson, $M_{BL}$ in \Tref{tab1},
computed at the vacuum of \Eref{vevs}, $\vev{M_{BL}}$. Such an
approach allows to find $\ck$ in terms of $\Mgut$ as follows
\beq \label{mgut}
\vev{M_{BL}}=g\vev{\sg}=\mgut\>\>\stackrel{(\ref{igc2})}{\Longrightarrow}\>\>\ck=g
^2\mP^2/\mgut^2.\eeq
It is beneficial that our scheme independently constrains the
allowed margin of $\mgut$ (and via it $M$) for given $g$ as we
show in \Sref{num}.


\paragraph{(b) Validity of the Effective Theory}

Despite the fact that the resulting $\ck$ in \Eref{mgut} is very
large, the effective theory is valid up to a high enough UV
cut-off scale $\Qef$ which can be derived by analyzing
\cite{actpal} the small-field behavior of our models in the EF --
cf. \cref{un0,riotto,un1,un2,un3,un4,un5,udemir}. Given that $\sg$
is canonically normalized we focus exclusively on $\Vhi$, see
\Eref{Vhi}, and expand it in terms of $\dph=\sg-M$ about the
vacuum in \Eref{vevs}. Neglecting irrelevant $\dn$-depended
contributions we arrive at the result
\beq\label{qef}\Vhi\simeq\frac{\ld^2}{64\ck^3\sg^2}
\lf1-3\frac{\dph}{\Qef}+\frac{25}{4}\frac{\dph^2}{\Qef^2}-11\frac{\dph^3}{\Qef^3}+
\frac{35}{2}\frac{\dph^4}{\Qef^4}+\cdots\rg~~\mbox{with}~~
\Qef=\frac{\mP}{\sqrt{\ca}},\eeq
for any $|\dn|\ll1$. Although $\Qef<\mP$, it is much larger than
$\Vhi(\sg)$ for the $\sg$ values during \fhi, assuring thereby
protection from dangerous loop-corrections. Indeed, as we show in
\Sref{num}, there is sizable parameter space consistent with the
condition
\beq \Vhi^{1/4}(\sg)\leq\Qef~~\mbox{for}~~\sg\leq1.
\label{vqef}\eeq
On the other hand, the fact that $\sgx\gg\Qef$ -- which causes, in
principle, concerns regarding corrections from non-renormalizable
terms associated with $\Qef$ -- does not invalidate our proposal,
since it is widely believed that dangerous loop-corrections depend
on the energy scale and not on the field values. Conservatively,
we impose $\sgx<1$ to avoid possible SUGRA corrections from higher
order terms. Indeed, terms of the form $(\bar\Phi\Phi)^\ell$ with
$\ell>1$ in \eqs{whi}{ktot} can not be prohibited by any symmetry
and so their harmlessness can be only ensured for this range of
$\sg$ values.

\subsection{Results}\label{res}

The delineation of the allowed parameter space of our model is the
scope of this section. We start with the derivation of some
approximate expressions in \Sref{ana} which assist us to interpret
the numerical results exhibited in \Sref{num}.

\subsubsection{Analytic Results}\label{ana}

Our analytic results can be obtained employing exclusively \Vhi\
in \Eref{Vhi}. The slow-roll parameters read
\beq
\label{srg}\eph=\frac{8\fw^2(1-\dn\fw)^2}{\ck^{4(1+\dn)}\sg^{2(5+4\dn)}}~~\mbox{and}~~
\ith=4\frac{5-3\ck\sg^2+\dn\fw(\ck\sg^2-4\dn\fw-9)}{\ck^{4(1+\dn)}\sg^{2(3+2\dn)}}\,.
\eeq
The termination of IPI is triggered by the violation of the $\ith$
criterion in \Eref{srcon} for $\sg=\sgf$ and it is largely
independent from the tiny $\dn$ values used in our work -- see
\Sref{num} below. Solving the equation $\ith(\sgf)=-1$ we find
\beq \sgf\simeq 2^{1/6}\re\frac{((d_{\rm
f}-5)^{2/3}+2^{4/3}\ck^{1/3})^{1/2}}{(d_{\rm
f}-5)^{1/6}\ck^{1/3}}~~\mbox{with}~~d_{\rm f}=\sqrt{25-16\ck}.
\label{sgf}\eeq
E.g., we obtain $0.045\lesssim\sgf\lesssim0.34$ for the allowed
regions in \Eref{res0} -- see \Sref{num} below.

We proceed our analysis below presenting separately our results
for $\dn=0$ and $\dn\neq0$.

\paragraph{(a) $\dn=0$ Case}\label{ana0}

In this case we reproduce the results exposed in \cref{actpal} for
$n=4$ and $m=1$. In particular, as we can verify a posteriori
$\sgf\ll\sgx$ and so, $\Ns$ can be calculated via \Eref{Nhi} as
follows
\beqs\beq \label{s*} \Ns\simeq\ck\sgx^4/16\>\Rightarrow\>
\sgx\simeq2(\Ns/\ck)^{1/4}.\eeq
Obviously, \fhi\ with \sub\ $\sg$'s can be attained if
\beq \label{fsub} \sgx\leq1~~\Rightarrow~~\ck\geq16\Ns\simeq800
\eeq\eeqs
for $\Ns\simeq50$. Therefore, large $\ck$'s are dictated as in the
metric formulation, but with larger magnitude. Nonetheless, the
bound above can be rephrased as an upper bound on
$\mgut\lesssim60~\YeV$ via \Eref{mgut} in excellent agreement with
our numerical results in \Eref{res0} -- see \Sref{num} below.

Replacing $\Vhi$ from \Eref{Vhi} in the left-hand side formula in
\Eref{Nhi} we find
\begin{equation} \As^{1/2}=\frac{\ \sgx^3}{32\pi}\sqrt{\frac{\ck}{3 +
3\ck \sgx^2}}
\>\>\Rightarrow\>\>\ld\simeq2\pi\sqrt{{6\As}}(\ck/\Ns)^{3/4}.
\label{lang} \eeq
Inserting, finally, $\sgx$ from \Eref{fsub} into \Eref{ns} we can
achieve the following expressions \cite{actpal}
\beq \label{gns}  \ns\simeq1-\frac{3}{2\Ns},~~
\as\simeq-\frac{3}{2\Ns^2}~~~\mbox{and}~~~
r\simeq\frac{2}{(\ck\Ns^{3})^{1/2}},\eeq
where $r$ is inversely proportional to $\ck^{1/2}$ and so well
below the value ($0.003$) obtained in metric formulation where $r$
is independent from $\ck$.

\paragraph{(b) $\dn\neq0$ Case}\label{ana1}

In this case $\Ns$ can be estimated again through \Eref{Nhi} with
result
\beq \Ns \simeq -\frac{\ck^{2\dn}\sgx^{2(1+2\dn)}}{8\dn(1+2\dn)}
\lf1+\frac{\ln(1-\dn\ck\sg^2)}{\dn\ck\sgx^2}\rg\label{Ngm}\eeq
which, obviously, can not be reduced to the one found for $\dn=0$
-- cf. \Eref{s*}. Due to the complicate functional form of $\Ns$
and the low $\dn$ values needed -- see \Sref{num} -- it is not
doable to solve \Eref{Ngm} and repeat the analysis of the previous
paragraph. Therefore, our last resort is the numerical computation
whose the results are presented in the following.

\subsubsection{Numerical Results}\label{num}


\begin{figure}[t]\vspace*{-.27in}
\hspace*{-.in}\begin{minipage}{75mm}
\epsfig{file=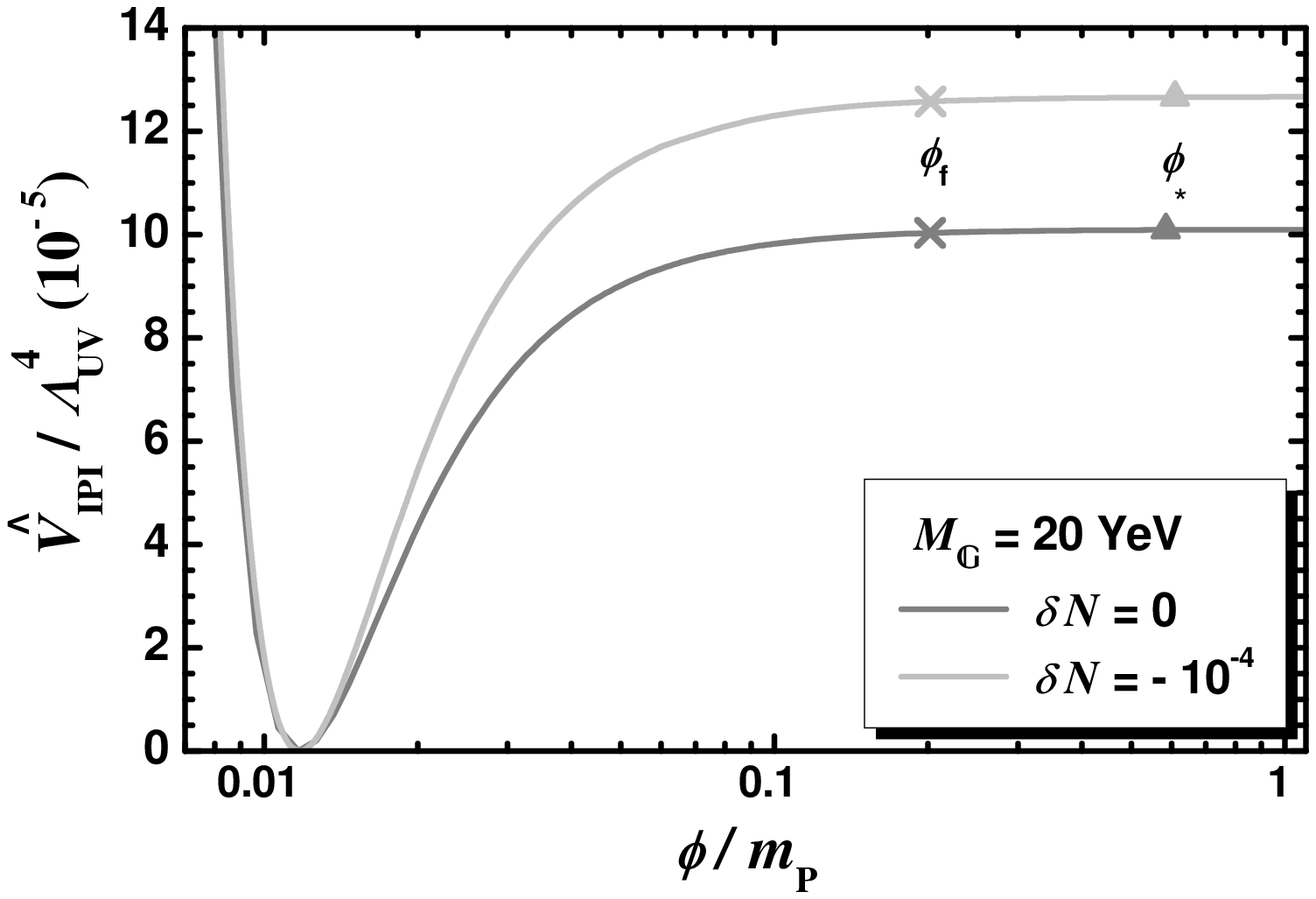,height=3.6in,angle=-90}
\end{minipage}
\hfill
\begin{minipage}{99mm}\renewcommand{\arraystretch}{1.2}
\vspace*{-.0in}\begin{center} {\small
\begin{tabular}{|c||c|c|}\hline
{\sc Model}&\multicolumn{2}{|c|}{$\dn$}\\\cline{2-3}
{\sc Parameters}&$0$&{$-10^{-4}$}\\ \hline\hline
$\ld$&$0.04$&$0.045$\\
$\sgx/\mP$&{$0.585$}&$0.61$\\
$\sgf/\mP$&{$0.2$}&$0.2$\\\hline
$\ns$&{$0.9719$}&$0.9739$\\
$r/10^{-5}$&{$6.1$}&$7.6$\\
$-\as/10^{-4}$&$5.3$&$5.6$ \\ \hline
\end{tabular}}
\end{center}\renewcommand{\arraystretch}{1.0}
\end{minipage}\vspace*{-.0in}
\hfill \caption{\sl\small Inflationary potential $\Vhi$ in units
of $\Qef$ as a function of $\sg$ for $\sg>0$, $g=0.7$,
$\mgut=20~\YeV$ and $\dn=0$ (dark gray line) or $\dn=10^{-4}$
(light gray line) -- the values of $\sgx$ and $\sgf$ are also
indicated. The relevant values of some input and output parameters
are listed in the Table -- recall that $1~\YeV=10^{15}~\GeV$.}
\label{fig1}\end{figure}


At a first glance, our model based on $W$ and $K$ in
\eqs{whi}{ktot} depend on the following parameters:
$$\ld,\>M,\>\ck,\>N\>\>\mbox{and}\>\>\nst.$$
Note that $N$ can be replaced by $\dn$ defined in \Eref{dndef}.
Taking into account, moreover, \eqs{igc2}{mgut} we can replace $M$
with $\ck$ and the last one with $\mgut$ for given $g$. Then $\ld$
and $\sgx$ can be determined by enforcing \Eref{prob} with
$\Trh\simeq10~\EeV$ -- which is consistent with the proposed
post-inflationary completion in \Sref{pfhi}. Possible variation
over 1-2 orders of magnitude may generate some minor modification
on $\Ns$. Moreover we fix $\nst=1$ throughout to assure that
$m_s>0$. Therefore, we are left with just two variables, for given
$g$, $\mgut$ and $N$ or $\dn$, which may be restricted by the
remaining constraints in \eqs{data}{vqef}.

To obtain a first insight for the shape of the inflationary
potential in our set-up we plot in \Fref{fig1} $\Vhi/\Qef^4$ in
\Eref{Vhi} versus $\sg$ for $\dn=0$ (gray line) or $\dn=-10^{-4}$
(light gray line). We take
\beq
g=0.7,~\mgut=20~\YeV,~\ck=7.25\cdot10^3~~\mbox{and}~~M=28.6~\YeV,\label{para}\eeq
where the two latter quantities are obtained consistently with
\eqs{mgut}{vevs} and we restore units hereafter in this Section
for presentation purposes. In both cases we fix $\Ns=52.5$ and
arrange the values of some related parameters in the Table of
\Fref{fig1}. We remark that for both $\dn$ values $\Vhi$
comfortably satisfies \Eref{vqef} and exhibits the same vacuum
state (since $\ck$ is the same in both cases). The position of the
inflationary plateau, however, is clearly elevated for $\dn<0$.
This means that the corresponding $r$ is larger \cite{rriotto}
since $\eph$ is larger. Indeed, we find that the $r$ value for
$\dn=0$ is less than its value for $\dn=-10^{-4}$. Moreover,
$\ith<0$ increases as $\dn$ decreases and so $\ns$ is lower for
$\dn=0$ as shown in the Table of \Fref{fig1}. Indeed, we find
$\ith=-0.014$ [$\ith=-0.013$] for $\dn=0$ [$\dn=-10^{-4}$].

\begin{figure}[!t]\vspace*{-.12in}
\hspace*{-.17in}
\begin{minipage}{8in}
\epsfig{file=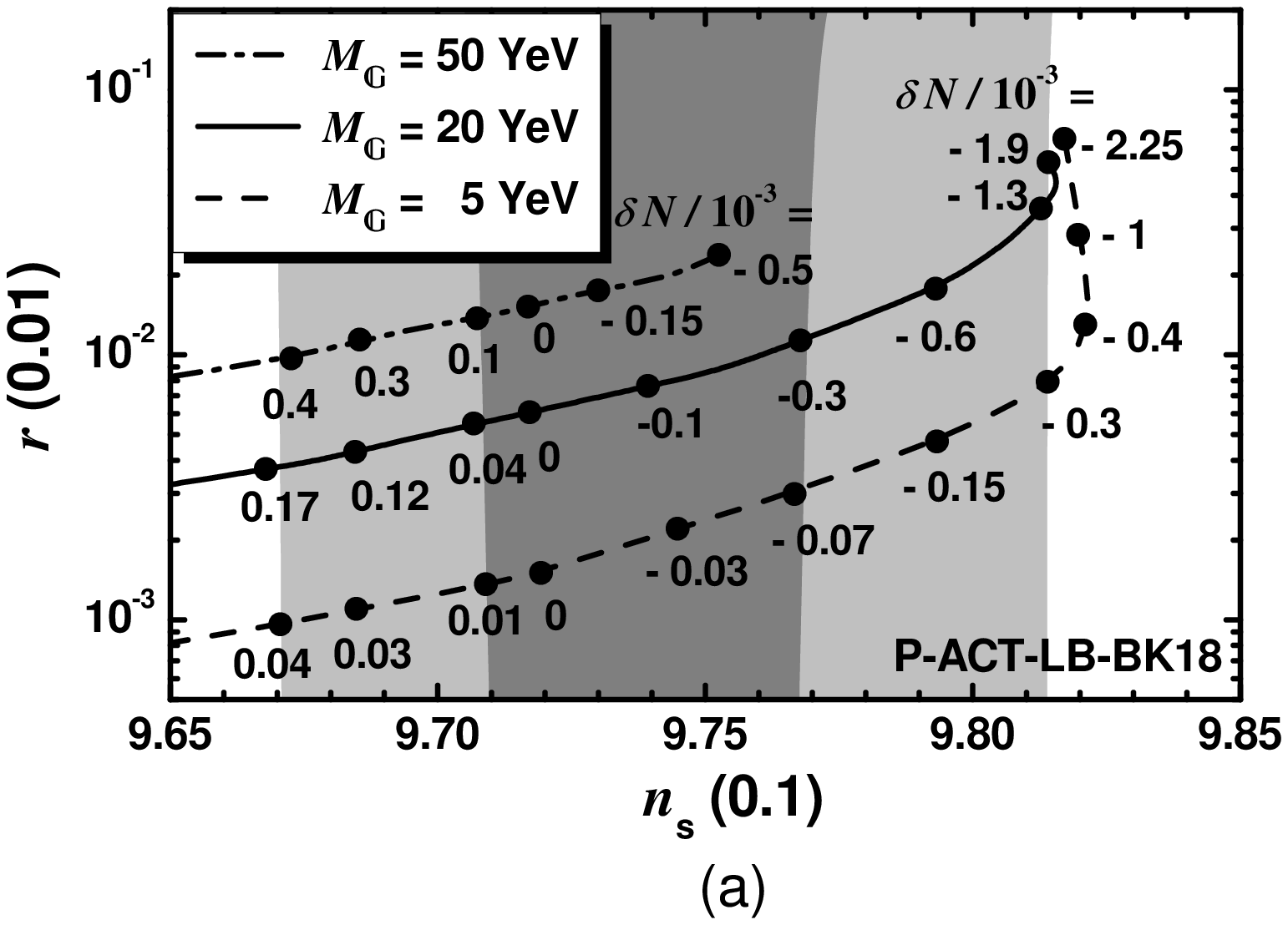,height=3.5in,angle=-90}
\hspace*{-1.cm}
\epsfig{file=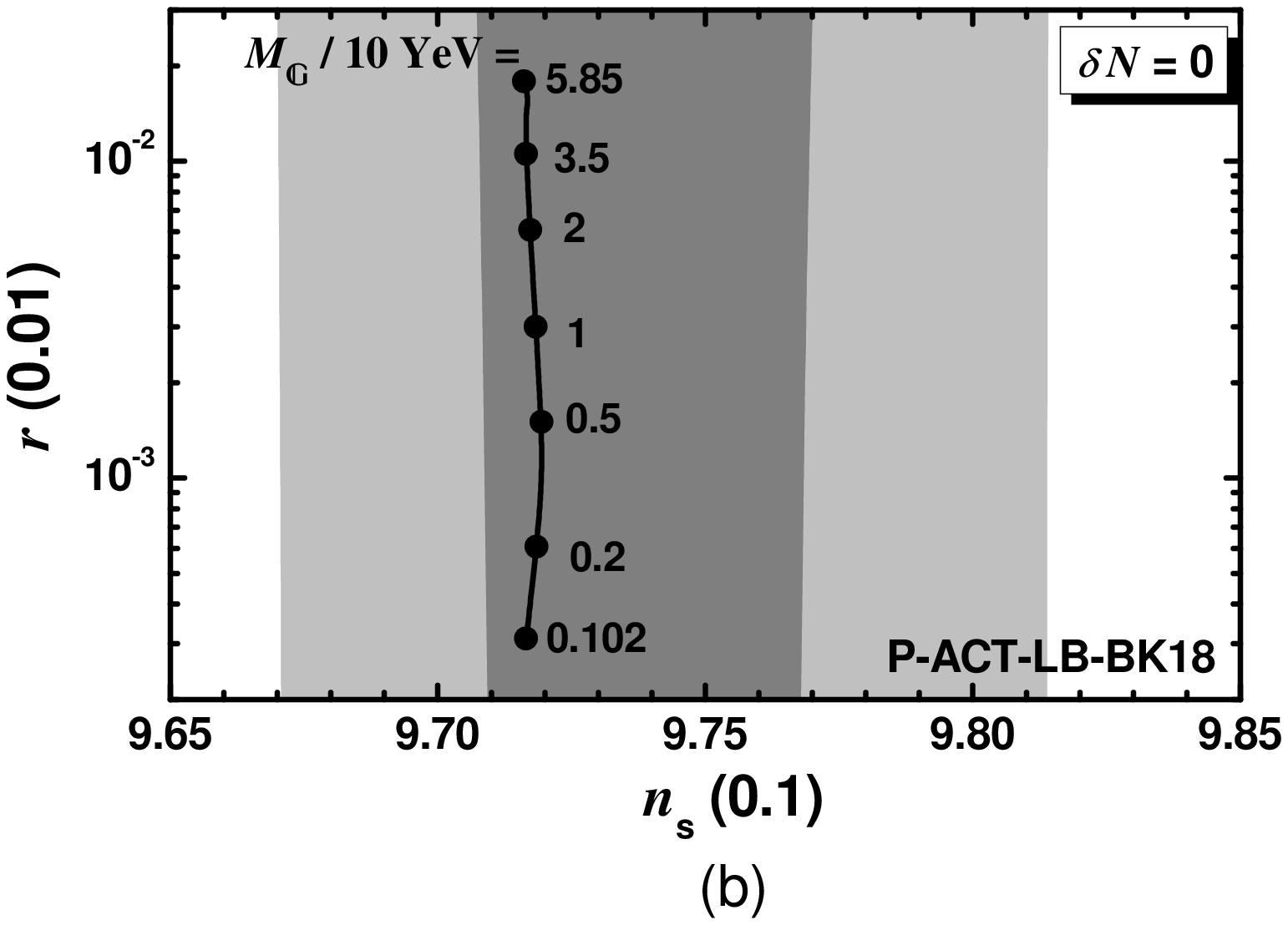,height=3.5in,angle=-90} \hfill
\end{minipage}
\hfill \caption{\sl\small Allowed curves in the $\ns-r$ plane for
$g=0.7$ and {\sffamily\ftn (a)} various $\mgut$'s shown in the
legend and varying $\dn$ as shown along the curves {\sffamily\ftn
(b)} $\dn=0$ and various $\mgut$'s shown along the curve. The
marginalized joint $68\%$ [$95\%$] regions from \actcf\ data are
depicted by the dark [light] shaded contours. }\label{fig2}
\end{figure}


The outputs of our numerical computation are first compared
against the \actc\ data \cite{actin} in the $\ns-r$ plane for
$g=0.7$ -- see \fref{fig2}. In both plots there we depict the
marginalized joint $68\%$ and $95\%$ regions from \actc\ data by
dark and light shaded contours respectively in the background. In
\sFref{fig2}{a} we also plot solid, dashed and dot-dashed lines
for $\mgut=5~\YeV, 20~\YeV$ and $50~\YeV$ correspondingly and show
the variation of $\dn$ along each line. We clearly see that $\ns$
and $r$ increase with $\mgut$ and as $\dn$ decreases. The various
lines terminate at large $\ns$ values, since the right inequality
in \Eref{vqef} is saturated. The variation of $\dn$ signals a
disturbing tuning due to its low magnitude. Indeed we obtain
$|dn|<10^{-3}$, i.e., three orders of magnitude lower than the
corresponding parameter ($n$) in \cref{ighi}. It is worth
emphasized, thought, that our proposal assures totally acceptable
results without any tuning, i.e., for $\dn=0$. This is clearly
illustrated in \sFref{fig2}{b}, where we display the allowed curve
in the $\ns-r$ plane for $N=2$ (i.e, $\dn=0$) and varying $\mgut$
along it. We see that the allowed curve lies well within the 68\%
c.l. of \actc\ data. Also the $\mgut$ values include the one
favored by the unification of the gauge coupling constants within
MSSM, $20~\YeV$. As anticipated below \Eref{mgut}, our scheme
restricts the allowed $\mgut$ (and $\ck$ via \Eref{mgut}) as
follows
\beq 1.02\lesssim\mgut/\YeV\lesssim58.5,~
2.8\cdot10^3\gtrsim\ck/10^3\gtrsim8.5~~\mbox{and}~~1.45\lesssim
M/\YeV\lesssim83.5\label{res0}\eeq
with $\Ns\simeq(52.1-52.9)$, $\as\simeq -5.3\cdot10^{-4}$ and
$r\simeq (0.3-18)\cdot10^{-5}$. For $\mgut$ values lower than the
lower bound on $\mgut$ in \Eref{res0} $\ld$ required by
\Eref{prob} exceeds the perturbative value $3.5$. On the other
hand, the upper bound on $\mgut$ in \Eref{res0} comes from the
requirement that $\sg$ has to be kept subplanckian. The obtained
$|\as|$'s remain negligibly small being, thereby, consistent with
\Eref{data}. The contribution of $\dV$ to $\Vhi$ in \Eref{vrc} can
be easily eliminated with a suitable selection of $\Ld$ as
mentioned in \Sref{fhi1}. E.g., imposing the condition
$\dV(\sgx)=0$ we find $\Ld_{\rm CW}\simeq(7.4-56)~\PeV$ for the
ranges in \Eref{res0}. Under these circumstances, our inflationary
predictions can be exclusively reproduced by using $\Vhi$ in
\Eref{Vhi}. If we vary $g$ in the range $0.5-0.9$ we obtain some
shift of the allowed $\mgut$ values whereas the resulting
observables remain intact. In particular, for $g=0.5$ we find
$0.725\lesssim\mgut/\YeV\lesssim41.5$ whereas for $g=0.9$ we find
$1.3\lesssim\mgut/\YeV\lesssim75$.

\begin{figure}[!t]\vspace*{-.27in}
\bec\includegraphics[width=60mm,angle=-90]{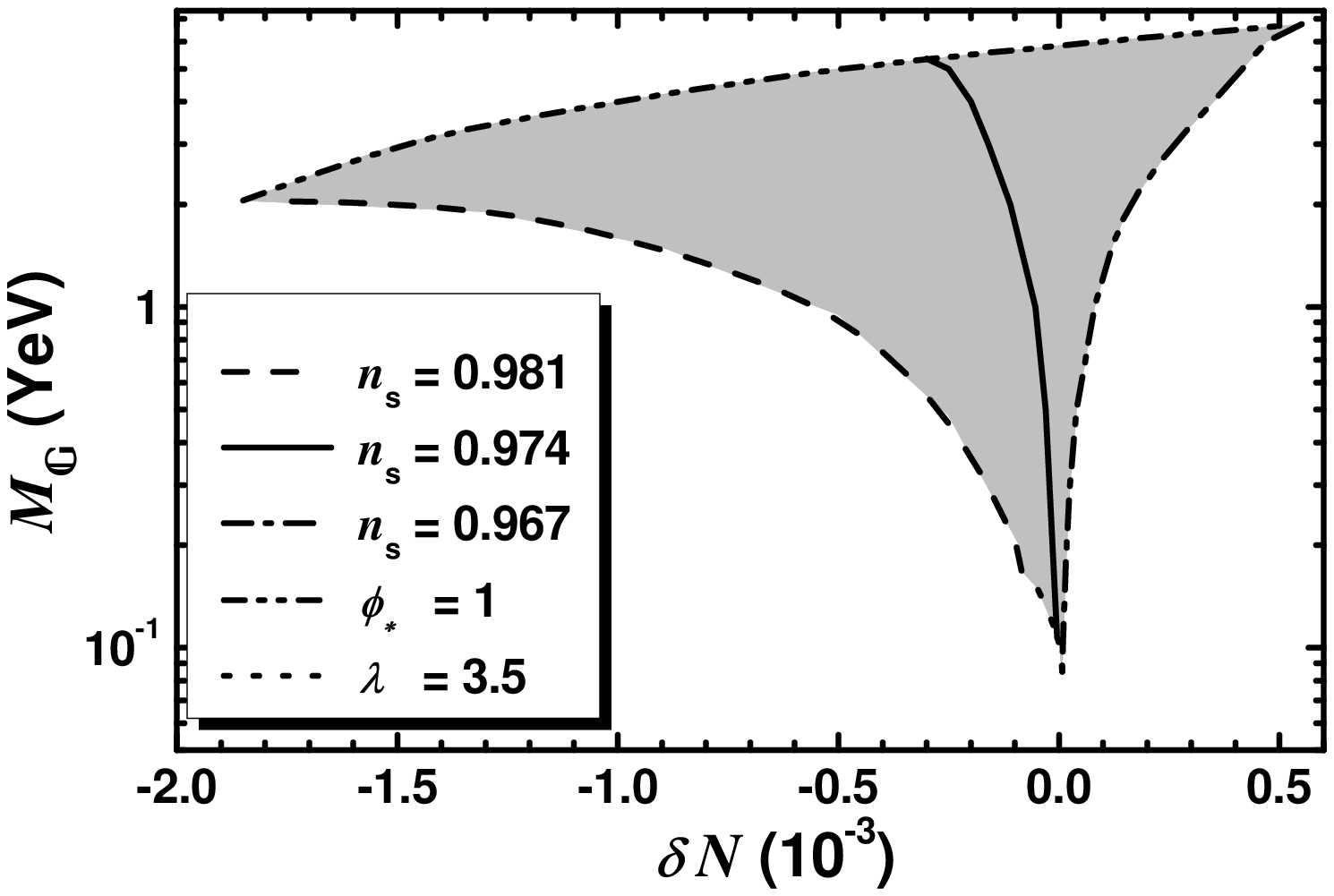}\eec
\hfill \caption{\sl\small  Allowed (shaded) region for $g=0.7$ as
determined by \eqsss{data}{prob}{mgut}{vqef} in the $\mgut-\dn$
plane. Along the solid line we fix $\ns=0.974$. The conventions
adopted for the various lines are also shown.}\label{fig3}
\end{figure}




The variation of the observables in \Fref{fig2} reveals that our
free parameters can be constrained in the $\dn-\mgut$ plane. The
relevant allowed region is shown in \Fref{fig3} for $g=0.7$. It is
bounded by {\ftn\sf (i)} the dashed and dot-dashed lines which
originate from the upper and the lower bounds on $\ns$ in
\Eref{data}; {\ftn\sf (ii)} the double dot-dashed line along which
$\sgx=\mP$ and {\ftn\sf (iii)} the dotted line along which
$\ld=3.5$ i.e., it acquires its maximal possible perturbative
value. Fixing $\ns$ to its central value in \Eref{data}, we obtain
the solid line with $\dn=-(5\cdot10^{-2}-3)\cdot10^{-4}$ and
values for the other parameters very close to those obtained for
$\dn=0$ in \Eref{res0}. Minor changes to the position of the
double dot-dashed and dotted curves are expected by varying $g$ in
the range $0.5-0.9$.


\section{Post-Inflationary Regime}
\label{pfhi}

A byproduct of our inflationary setting is that it assists us to
understand the origin of the $\mu$ term of MSSM, as we show in
\Sref{pfhi1}, consistently with the generation of the observable
baryon asymmetry -- see \Sref{pfhi2} --, if we embed it in a $B-L$
extension of MSSM as detailed in \Sref{pfhi0}. Our final results
are exposed in \Sref{respost}.

\subsection{Set-up}\label{pfhi0}

To achieve the aforementioned $B-L$ extension of MSSM we promote
to gauge the pre-existing global $U(1)_{B-L}$ -- cf. \cref{nmBL,
tmhi}. The supplementary superpotential terms which are relevant
for our analysis read
\beqs\beq\label{dW} \dW=\ld_{\mu} S\hu\hd\ +\  h_{ijN} \sni L_j
\hu +\lrh[i]\phcb N^{c2}_i.\eeq
On the other hand, $K_2$ in \Eref{kb} is to be replaced by
\beq
K_2=\nst\ln\lf1+\mbox{$\sum_{a=1}^{5}$}|X_\al|^2/\nst\rg\label{dK}\eeq\eeqs
with~~$X_\al=S, \wtilde N^{c}_i, \wtilde L_j, \hu$ and $\hd$ which
parameterizes the compact \Km\ $SU(15)/U(1)$ with constant scalar
curvature $14\cdot15/\nst=210/\nst$. Here $\hu$ and $\hd$ are the
electroweak Higgs superfields and $\ssni$ and $\wtilde L_j$ are
the scalar superpartners of the right-handed neutrinos $\sni$ and
left-handed leptons $L_j$ respectively with $i=1,2,3$ -- we work
in the basis where $\lrh[i]$ is diagonal, real and positive. The
charge assignments of $X^\al$ besides $S$ are
\beq (B-L)(L_i, N^{c}_i, \hu, \hd)=(-1,1,0,0)~~\mbox{and}~~R(L_i,
N^{c}_i, \hu, \hd)=(0,1,0,0).\eeq
We assume that $X^\al$ are stabilized at zero during \fhi\ and so
the inflationary trajectory in \Eref{inftr} has to be supplemented
by the condition
\beq\label{inftr1}
\vevi{\hu}=\vevi{\hd}=\vevi{\ssni}=\vevi{\wtilde L_j}=0.\eeq
The consistency of our assumption can be verified by checking the
stability of this path. To this end we parameterize the complex
fields $X^\al$ above as $S$ in \Eref{hpar}, i.e.,
\beq X^\al= ({x^\al +i\bar x^\al})/{\sqrt{2}}.\eeq
The relevant masses squared are listed in \Tref{tab2}, where we
see that $m_{i\wtilde \nu^c}^2>0$, $m_{i\wtilde l}^2>0$ and
$m^2_{h+}>0$ for any $\sgf<\sg<1$ -- here $m_{h\pm}$ are the
eigenvalues associated with the eigenstates of Higgs doublets
\beq h_\pm=(h_u\pm{h_d})/\sqrt{2}\>\>\>\mbox{and}\>\>\> {\bar
h}_\pm=({\bar h}_u\pm{\bar h}_d)/\sqrt{2}. \eeq
Note that $m_s$ remains as shown in \Tref{tab1}. On the other
hand, the positivity of $m^2_{h-}$ dictates the establishment of
the inequality -- cf. \cref{tmhi,phi}:
\beq
\lm<\ld(1+\nst)(\ck\sgf^2-1)/4\nst\ck~~\Rightarrow~~\lm/\ld<(1+1/\nst)\sgf^2/4,\label{lmb}
\eeq
where $\sgf$ is given by \Eref{sgf}. E.g., for the ranges of
parameters in \Eref{res0} we obtain
$0.001\lesssim\lm/\ld\lesssim0.06$. Due to the extra contributions
of \Tref{tab2} the total radiative correction to $\Vhi$ in
\Eref{Vhi} now becomes
\beq \Delta V_{\rm tot}=\dV+ \Delta V_{\rm pI}\eeq
where $\dV$ is given by \Eref{vrc}, whereas the latter term takes
the form
\beq \label{dV2}\Delta V_{\rm pI}=\frac{1}{64\pi^2}\lf\sum_\al
{{\rm N}_\al} m_\al^4\ln \frac{m_\al^2}{\Ld_{\rm CW}^2}-
6M_{iN^c}^4\ln \frac{M_{iN^c}^2}{\Ld_{\rm
CW}^2}\rg~~\mbox{where}~~\bcs \al&=\{h\pm,i\wtilde \nu^c,i\wtilde
l\}\\ {\rm N}_\al&=\{8,6,12\}\ecs.\eeq
As a consequence, $\Ld_{\rm CW}$ should be readjusted if we impose
the condition $\Delta V_{\rm tot}(\sgx)=0$. E.g., for the ranges
of parameters in \Eref{res0} we find $\Ld_{\rm
CW}\simeq(0.77-5.85)~\ZeV$ -- hereafter we restore units.


\renewcommand{\arraystretch}{1.4}
\begin{table}[!t]
\bec\begin{tabular}{|c|c|c|c|c|}\hline
{\sc Fields}&{\sc Eigenstates} & \multicolumn{3}{c|}{\sc Masses
Squared}\\\hline\hline
26 Real & $h_{\pm},{\bar h}_{\pm}$ &
$m_{h\pm}^2$&\multicolumn{2}{c|}{$3\lf1+1/\nst\pm{4\lm\ck}/{\ld\fw}\rg\Hhi^2$}\\
Scalars  & $\wtilde \nu^c_{i}, \bar{\wtilde\nu}^c_{i}$ & $
m_{i\wtilde
\nu^c}^2$&\multicolumn{2}{c|}{$3\lf1+1/\nst+16\ld^2_{iN^c
}/\ld^2\sg^2\rg\Hhi^2$} \\
& $\wtilde l_{i}, \bar{\wtilde l}_{i}$ & $ m_{i\wtilde
l}^2$&\multicolumn{2}{c|}{$3(1+1/\nst)\Hhi^2$} \\\hline
$3$ Weyl Spinors&{$N_i^c$}& {$
M_{{iN^c}}^2$}&\multicolumn{2}{c|}{{$48\ld^2_{iN^c}\ck^2\sg^2\Hhi^2/\ld^2\fw^2$}}\\
\hline
\end{tabular}\eec
\caption{\sl\small Mass-squared spectrum of the non-inflaton
sector along the path in
\eqs{inftr}{inftr1}.}\label{tab2}\end{table}
\renewcommand{\arraystretch}{1.}

\subsection{Generation of the $\mu$ Term of MSSM}\label{pfhi1}

The contributions from the soft SUSY-breaking terms, although
negligible during IPI -- since these are much smaller than
$\sg\sim\mP$ -- may shift slightly $\vev{S}$ from zero in
\Eref{vevs}. Indeed, the relevant potential terms are
\beq V_{\rm soft}= \lf\ld A_\ld S \phcb\phc+\lm A_\mu S \hu\hd +
\ld_{iN^c} A_{iN^c}\phc \widetilde N^{c2}_i- {\rm a}_{S}S\ld M^2/4
+ {\rm h. c.}\rg+ m_{\al}^2\left|X^\al\right|^2, \label{Vsoft}
\eeq
where $m_{\al}, A_\ld, A_\mu, A_{iN^c}$ and $\aS$ are soft
SUSY-breaking mass parameters.  Rotating $S$ in the real axis by
an appropriate $R$ transformation, choosing conveniently the
phases of $\Ald$ and $\aS$ so as the total low energy potential
$V_{\rm tot}=V_{\rm SUSY}+V_{\rm soft}$ to be minimized -- see
\Eref{VF} -- and substituting $\hu=\hd=0$ and $\phc=\phcb=M/2$
from \Eref{vevs} we get
\beqs\beq \vev{V_{\rm tot}(S)}= \frac12\ld^2{M^2S^2} -\ld M^2S\am
\mgr ~~\mbox{with}~~~\am=\lf|A_\ld| + |{\rm a}_{S}|\rg/2\mgr,
\label{Vol} \eeq
where $\mgr$ is the $\Gr$ (gravitino) mass and $\am>0$ is a
parameter of order unity which parameterizes our ignorance for the
dependence of $|A_\ld|$ and $|{\rm a}_{S}|$ on $\mgr$. We also
take into account that $m_S\ll M$.  The extermination condition
for $\vev{V_{\rm tot}(S)}$ w.r.t $S$ leads to a non vanishing
$\vev{S}$ as follows
\beq \label{vevS}{d}\vev{V_{\rm tot}(S)}/{d S}
=0~~\Rightarrow~~~\vev{S}\simeq \am\mgr/{\ld},\eeq\eeqs
The extremum above is a global minimum since ${d^2} \vev{V_{\rm
tot}(S)}/{d S^2}=\ld^2M^2>0$. { The SUSY breaking effects,
considered in \Eref{Vsoft}, explicitly break $U(1)_R$ to the
subgroup $\mathbb{Z}_2^{R}$, which remains unbroken by $\vev{S}$
in \Eref{vevS} and so no disastrous domain walls are formed.}

The generated $\mu$ parameter from the first term in \Eref{dW} is
\beq\mu =\lm \vev{S} \simeq {\lm}\am\mgr/\ld.\label{mu}\eeq
This results reveals that $\mu/\mgr$ follows the behavior of the
ratio $\lm/\ld$ shown below \Eref{lmb}. The emerged hierarchy
between $\mu$ and $\mgr$ is mildly stronger than that obtained in
\cref{nmBL,R2r,ighi,ighic,phi}, rendering our present setting
apparently distinguishable from those. Namely, our present result
hints at split SUSY \cite{split, split1}, where the soft SUSY
scalar mass parameters -- and therefore, the SUSY mass scale,
$\mss$ -- are expected to lie at an intermediate mass scale
whereas $\mu$ and the gaugino masses can be kept of the order of
\TeV. As a consequence, the candidacy of the \emph{lightest SUSY
particle} ({\sf\ftn LSP}) -- which is normally the lightest
neutralino, $\nta$ -- as \emph{cold dark matter} ({\sf\ftn CDM})
particle remains possible and the gauge coupling unification is
still valid for $g$ and $\mgut$ values similar to those used in
the conventional MSSM.

Given that the LEP bound on the mass of charginos \cite{lep}
dictates $\mu>0.1~\TeV$, \Eref{mu} consistently with
\eqs{lang}{mgut} favors $\mgr>1~\PeV$. On the other hand, if
$\mss$ is identified with $\mgr$, our proposal can be probed via
the measured value of the Higgs boson mass \cite{lhc}. Within
split SUSY, updated analysis requires \cite{strumia}
\beq 10~\TeV\lesssim\mgr\lesssim60~\PeV,\label{mssb} \eeq
for degenerate sparticle spectrum and varying $\tan\beta$ and stop
mixing. The lower bound above is moved up to $20~\PeV$ for low
$\tan\beta$ values and minimal stop mixing. The upper bound above
can be moved up to $100~\PeV$ for scalar mass parameters larger or
smaller than \mgr\ by a factor of $3$. Note that the upper bound
on \Eref{mssb} automatically renders the gluino with mass of order
$1~\TeV$ cosmologically safe \cite{gluino}.

\subsection{Non-Thermal Leptogenesis after IPI}\label{pfhi2}

Besides the generation of the $\mu$ term, $\dW$ in \Eref{dW}
allows for the implementation of nTL via the direct decay of the
inflaton into $\rhni$ -- see \Sref{lept0}. For the success of this
process, however, one has to produce the correct baryon asymmetry
-- see \Sref{lept1} -- and circumvent the $\Gr$ overproduction
problem as discussed in \Sref{lept2}.

\subsubsection{Inflaton Mass \& Decay}\label{lept0}

Soon after the end of IPI, we obtain a very short period of OI --
see Appendix~\ref{app} -- which is followed by the usual phase of
damped oscillations abound the minimum in \Eref{vevs} where the
(canonically normalized) inflaton $\dph=\phi-M$ acquires mass
\beq \label{msn} %
\msn=\left\langle\Ve_{\rm IPI,\sg\sg}\right\rangle^{1/2}= {\ld
\mP}/{\sqrt{2\ck}},\eeq
which ranges in the interval $(4.76-36)\cdot10^{-1}~\YeV$ with
decreasing $\mgut$ (or increasing $\ck$). Note that it does not
depend on $N$ (and $\dn$) thanks to the canonical normalization of
$\dph$. This oscillatory period terminates at the moment of
reheating where the temperature of the universe acquires the value
\cite{phi}
\beq\Trh=
\left({72/5\pi^2g_{*}}\right)^{1/4}\lf\Gsn\mP\rg^{1/2}\>\>\>\mbox{with}\>\>\>\Gsn=\GNsn+\Ghsn+\Gysn,\label{Trh}\eeq
where $g_{*}=228.75$ counts the MSSM effective number of
relativistic degrees of freedom and we take into account the
following decay widths
\beqs\bea \GNsn&=&\frac{g_{iN^c}^2}{16\pi}\msn\lf1-{4\mrh[
i]^2}/{\msn^2}\rg^{3/2}\>\>\mbox{with}\>\>\>
g_{iN^c}=(N-1){\ld_{iN^c}};\label{dec1}\\
\Ghsn&=&\frac{2}{8\pi}g_{H}^2\msn\>\>\>\>\mbox{with}\>\>\>\>
g_{H}={\lm}/{\sqrt{2}}; \label{dec2}\\
\Gysn&=&\frac{14
g_y^2}{512\pi^3}\frac{\msn^3}{\mP^2}\>\>\>\>\mbox{with}\>\>\>\>g_y=y_{3}N\ck^{1/2}.\label{dec3}\eea\eeqs
Here $y_{3}=h_{t,b,\tau}(\msn)\simeq0.5$ with $h_t, h_b$ and
$h_\tau$ being the Yukawa coupling constants -- we assume that
diagonalization has been performed in the generation space.  The
various decay widths above arise from the respective lagrangian
terms
\beqs\bea {\cal L}_{\dph\to \sni\sni}&=&
-\frac12e^{K/2\mP^2}\dW_{,N_i^cN^c_i}\sni\sni\ +\ {\rm
h.c.}=g_{iN^c} \dph\ \lf\sni\sni\ +\ {\rm h.c.}\rg +\cdots,\\
{\cal
L}_{\dph\to\hu\hd}&=&-e^{K/\mP^2}K^{SS^*}\left|\dW_{,S}\right|^2
=-g_{H} \msn\dph \lf H_u^*H_d^*\ +\ {\rm h.c.}\rg+\cdots, \\
{\cal L}_{\dph\to XYZ}&=&g_y(\dph/\mP)\lf
X\psi_{Y}\psi_{Z}+Y\psi_{X}\psi_{Z}+ Z\psi_{X}\psi_{Y}+{\rm
h.c.}\rg, \label{lint} \eea\eeqs
describing $\dph$ decay into a pair of $N^c_{j}$ with masses
$\mrh[j]=\ld_{jN^c}M$, $\hu$ and $\hd$ and three MSSM
(s)-particles $X, Y$ and $Z$, respectively. We remark that $g_y$
is proportional to $\ck^{1/2}$ and therefore it is rather enhanced
due to the large $\ck$ values employed in our scheme -- see, e.g.,
\Eref{res0}.

\subsubsection{Baryon-Number Abundance}\label{lept1}

For $\Trh<\mrh[i]$, the out-of-equilibrium decay of $N^c_{i}$
generates a lepton-number asymmetry (per $N^c_{i}$ decay),
$\ve_i$. The resulting lepton-number asymmetry is partially
converted through sphaleron effects into a yield of the observed
baryon asymmetry of the universe
\beq Y_B=-0.35\cdot{5\over4}{\Trh\over\msn}\mbox{$\sum_i$}
{\GNsn\over\Gsn}\ve_i,\label{Yb}\eeq
where the quantity $\ve_i$ can be expressed in terms of the Dirac
masses of $\nu_i$, $\mD[i]$, arising from the second term of
\Eref{dW} -- see \cref{nmBL}. Moreover, employing the type I
seesaw formula we can then obtain the light-neutrino mass matrix
$m_\nu$ in terms of $\mD[i]$ and $\mrh[i]$. As a consequence, nTL
can be nicely linked to low energy neutrino data. We take as
inputs the recently updated best-fit values \cite{valle} -- cf.
\cref{nmBL} -- on the neutrino oscillation parameters including
IceCube IC24 with Super Kamiokande (SK) atmospheric data. We
consider only the scheme of \emph{normal ordered} \emph{neutrino
masses}, $\mn[i]$'s, which can become consistent with the upper
bound on the sum of $\mn[i]$'s
\beq\mbox{$\sum_i$} \mn[i]\leq0.082~{\eV},\label{sumnu}\eeq
induced from {\sf\ftn P-ACT-LB} data \cite{act} at 95\% c.l.
Namely, we take \cite{valle} $\Delta
m^2_{21}=7.49\cdot10^{-5}~{\rm eV}^2$ and $\Delta
m^2_{31}=2.513\cdot10^{-3}~{\rm eV}^2$ for the mass-squared
differences, $\sin^2\theta_{12}=0.308$,
$\sin^2\theta_{13}=0.02215$ and $\sin^2\theta_{23}=0.47$ for the
mixing angles, and $\delta=1.1788\pi$ for the CP-violating Dirac
phase.

The validity of \Eref{Yb} requires that the $\dph$ decay into a
pair of $\sni$'s is kinematically allowed for at least one species
of the $\sni$'s and also that there is no erasure of the produced
lepton yield due to $N^c_1$ mediated inverse decays and $\Delta
L=1$ scatterings. These prerequisites are ensured if we impose
\beq\label{kin}  10\Trh \lesssim\mrh[1] \leq\msn/2. \eeq
Finally, \Eref{Yb} has to reproduce the observational result
\cite{act}
\beq
Y_B=\lf8.75\pm0.085\rg\cdot10^{-11}~~\Rightarrow~~8.665\lesssim
Y_B/10^{-11}\lesssim8.835.\label{bdata}\eeq

\subsubsection{Gravitino Abundance}\label{lept2}

The required $\Trh$ in \Eref{Yb} must be compatible with
constraints on the $\Gr$ abundance, $\Yg$, at the onset of
\emph{nucleosynthesis} ({\ftn\sf BBN}), which is estimated to be
\cite{brand,kohri,grspanos}
\beq\label{Ygr} \Yg\simeq 1.9\cdot10^{-13}\ \Trh/\EeV,\eeq
where we assume that $\Gr$ decays with a tiny hadronic branching
ratio and take into account only thermal production of $\Gr$, and
assume that $\Gr$ is much heavier than the MSSM gauginos.
Non-thermal contributions to $\Yg$ \cite{idec} are also possible
but strongly dependent on the mechanism of soft SUSY breaking. It
is notable, though, that these contributions to $\Yg$ in models
with stabilizer field, as in our case, are significantly
suppressed compared to the thermal ones \cite{grNew}.

On the other hand, $\Yg$  is bounded from above in order to avoid
spoiling the success of the BBN. The corresponding bounds result
to upper bounds on $\Trh$ -- cf. \cref{phi} -- which are violated
in our case since the achievement of the correct $Y_B$ requires
$\Trh>10~\EeV$ as we see in \Tref{tab3}. We are obliged,
therefore, to adopt one of the following two alternative
scenarios:

\paragraph{(a) Short-lived $\Gr$} In this scenario, $\Gr$ decays before
BBN -- but after LSP freeze-out -- and so, the BBN bounds on
$\Trh$ no longer apply. To accommodate this situation we compute
the $\Gr$ decay temperature $\Tgr$ obtained by the condition
$H(\Tgr)\simeq\Ggr$ where \Ggr\ is the decay width of $\Gr$. We
find
\beq\Tgr= \left({90/\pi^2g_{\rm
SM*}}\right)^{1/4}\lf\Ggr\mP\rg^{1/2}\>\>\>\mbox{where}\>\>\>\Ggr=
\frac{1}{32\pi}\lf\frac13 N_{\rm ch}+N_{\rm
gs}\rg\frac{\mgr^3}{\mP^2}\label{Tgr}\eeq
is estimated taking the large $\Gr$ mass limit with $N_{\rm
ch}=49$ and $N_{\rm gs} = 12$ being \cite{grdecay} the number of
chiral superfields (neglecting $\rhni$'s) and the number of
gauginos in the MSSM respectively. { Also $g_{\rm SM*}=106.75$ is
the effective number of the relativistic degrees of freedom of the
Standard Model.} Requiring that $\Tgr$ is larger than the
temperature at the commencement of BBN, $\Tns\simeq2~\MeV$ we
obtain the lower bound on $\mgr$, $\mgr>50~\TeV$. On the other
hand, the abundance of the neutralino, $\nta$, LSP obtained by the
$\Gr$ decay is estimated to be
\beq \Omx=2.75\cdot10^{11}\Yg\mx/1~\TeV, \label{omx}\eeq
%
where $\Yg$ is given by \Eref{Ygr} -- $\Omega$ here should not be
confused with the frame function in \Eref{weyl}. The resulting
$\Omx$ should be consistent with upper bound from the CDM
considerations \cite{act} which provides us with an upper bound on
$\mx$ as follows
\beq
\Omx\lesssim0.12~~\stackrel{(\ref{omx})}{\Longrightarrow}~~\mx/1~\TeV\gtrsim2.3\lf1~\EeV/\Trh\rg.\label{mxb}\eeq
According to the updated results in \cref{xbound} a lower bound on
$\mx$ can be inferred by imposing a number of
cosmo-phenomenological requirements which can be translated to an
upper bound on $\Trh$ via \Eref{mxb}. In particular,
\beq
\mx\gtrsim40~\GeV~~\stackrel{(\ref{mxb})}{\Longrightarrow}~~\Trh\lesssim57~\EeV,
\label{mxmicro}\eeq
which is slightly higher than the upper bounds on $\Trh$ obtained
in the case of the long-lived $\Gr$ \cite{kohri} --
cf.~\cref{nmBL, igi, phi}. This relaxation is not enough for our
purposes as we see in \Sref{respost}.


\paragraph{(b) Very short-lived $\Gr$} This scenario allows for
$\Trh$ values higher than that in \Eref{mxmicro}. According to it,
$\Gr$ decays before the freeze-out process of $\nta$ LSP which
takes place for temperature $\Tf=\xf\mx>\Tns$ with
$\xf=1/25-1/20$. To achieve this we demand
\beq
\Tgr>\Tf~~\stackrel{(\ref{Tgr})}{\Longrightarrow}~~\mgr>2^{3/2}\lf\frac{g_{\rm
SM*}}{5}\rg^{1/6}\lf\frac{\mP(\pi\xf\mx)^2}{N_{\rm ch}+3N_{\rm
gs}}\rg^{1/3}, \label{T32f}\eeq
which can be translated numerically as
\beq \mgr> 3.1~\PeV (\xf\mx/1~\GeV)^{2/3}. \label{m32f}\eeq
E.g., for $\mx=1~\TeV$ we obtain $\mgr>36~\PeV$ [$\mgr>42~\PeV$]
with $\xf=1/25$ [$\xf=1/20$] which lies inside the region of
\Eref{mssb}, whose the upper bound requires $\mx\lesssim2.15~\TeV$
[$\mx\lesssim1.7~\TeV$] for $\xf=1/25$ [$\xf=1/20$].

\subsection{Results}\label{respost}

\renewcommand{\arraystretch}{1.25}
\begin{table}[!t]
\bec\begin{tabular}{|c||c|c|c|c|c|c|c|c|}\hline
{\sc Parameters} &  \multicolumn{8}{c|}{\sc Cases}\\\cline{2-9}
&A&B&C&D&E&F&G&H\\ \hline  \hline
\multicolumn{9}{|c|}{\sc Low Scale Parameters}\\\hline
$\mn[1]/0.001~\eV$&$0.1$&$0.79$&$0.869$&$1$ & $1$& $2$&$9.9$ & $10$\\
$\mn[2]/0.001~\eV$&$8.6$&$8.7$&$8.7$&$8.7$ & $8.7$& $8.8$ &$13.1$& $13.2$\\
$\mn[3]/0.01~\eV$&$5$&$5$&$5$&$5$ & $5$&$5$ &$5.1$& $5.1$\\\hline
$\sum_i\mn[i]/0.01~\eV$&$5.88$&$5.96$&$5.98 $&$5.98$&$5.9$ &$6.1$&
$7.41$ &$7.43$\\ \hline
$\varphi_1$&$\pi/20$&$0$&$-\pi/3$&$0$ & $\pi/7$&$0$&$\pi/10$& $0$\\
$\varphi_2$&$0$&$-\pi$&$2\pi$ &$\pi/25$& $-2\pi$&$-\pi/15$ &
$0$&$0$\\\hline
\multicolumn{9}{|c|}{\sc Leptogenesis-Scale Parameters}\\\hline
$\mD[1]/1~\GeV$&$93.8$&$79.81$&$86.3$&$89.6$ & $11.1$&$77$ &$116$& $120.1$\\
$\mD[2]/1~\GeV$&$70.1$&$60$&$95.98$&$89.6$ & $130.67$&$120.44$ &$80.595$& $121$\\
$\mD[3]/1~\GeV$&$169.46$&$111$&$100$&$90$ & $160$&$200$ &
$99.9$&$141.38$\\\hline
$\mrh[1]/0.1~\YeV$&$1.63$&$1.34$&$1.9 $&$1.61$ & $4.06$&$4.05$ &$1.68$& $3.3$\\
$\mrh[2]/1~\YeV$&$1.4$&$0.88$&$1$&$0.2$ & $1.9$&$1.17$ &$0.69$& $1.8$\\
$\mrh[3]/1~\YeV$&$126$&$11.2$&$8.45$&$8$ &
$15.8$&$8.12$&$1.8$&$1.6$\\\hline
\multicolumn{9}{|c|}{\sc Branching Ratio for $\dph$ Decay into
$\rhna$'s}\\\hline
$\Gamma_{\dph\to
N_1^c}/\Gsn~(\%)$&$41$&$33.9$&$46.1$&$40.6$&$7.7\cdot10^{-3}$&$0.29$
&$38$&$42$\\\hline
\multicolumn{9}{|c|}{\sc Resulting $\Trh$}\\\hline
$\Trh/10~\ZeV$&$1.4$&$1.3$&$1.44$&$1.37$&$1.06$&$1.06$
&$1.46$&$1.39$\\\hline
\multicolumn{9}{|c|}{\sc Resulting $B$-Yield }\\\hline
$Y_B/10^{-11}$&$8.83$&$8.83$&$8.75$&$8.71$ & $8.76$&$8.83$
&$8.83$&$8.82$\\\hline %
\end{tabular}\eec
\hfill \caption[]{\sl\small  Parameters yielding the correct $Y_B$
in \Eref{bdata} for various neutrino masses. We take $g=0.7$,
$\mgut=20~\YeV$, $N=2\nst=2$, $\mgr=40~\PeV$ and $\lm=10^{-6}$ --
recall that $1~\ZeV=10^{-3}~\YeV=10^{12}~\GeV$.} \label{tab3}
\end{table}

The numerical implementation of our post-inflationary setting can
be processed following the bottom-up approach detailed in
\cref{nmBL}. Namely, we find the $\mrh[i]$'s using as inputs the
$\mD[i]$'s, $\mn[1]$, the two Majorana phases $\varphi_1$ and
$\varphi_2$ of the PMNS matrix, and the best-fit values of the
neutrino oscillation parameters, mentioned in \Sref{pfhi2}. Since
the gauge group adopted here, $\Gbl$, no specific relation is
predicted between the Yukawa couplings constants $h_{iN}$ entering
the second term of \Eref{dW} and the other Yukawa couplings in the
MSSM. As a consequence, the $\mD[i]$'s are free parameters.
However, for the sake of comparison, for cases B, C, D and G we
take $\mD[3]=m_t\simeq100~\GeV$, where $m_t$ denotes the mass of
the top quark estimated at a scale close to $\msn$ in \Eref{msn},
i.e., $0.1~\YeV$.

Throughout our computation we take the values of parameters in
\Eref{para} and $\dn=0$ -- see Table in \Fref{fig1}. For these
inputs we obtain $\msn=0.8~\YeV$, independently of $\dn$, and
$\Ld_{\rm CW}\simeq3.4~\ZeV$. Also, we select $\nst=1$,
$\mu=1~\TeV$ (with soft SUSY-breaking masses for the gauginos of
the same order of magnitude) and $\mgr=40~\PeV$ which fulfils
\Eref{m32f}. As a result we have $\lm=10^{-6}$ which is consistent
with \Eref{lmb}. Finally, we take $y=0.5$ in \Eref{dec3}, which is
a typical value encountered for various MSSM settings.

In \Tref{tab3} we present eight benchmark points with increasing
$\mn[1]$'s, where all the restrictions of \Sref{pfhi2}, together
with those from \eqsss{data}{prob}{mgut}{vqef}, can be met. From
our computation, we remark the following:

\begin{itemize}

\item[{\sf\small (i)}] $\Ghsn$ turns out to be much less than
$\Gysn$ and $\GNsn$ which yield the dominant contributions into
$\Gsn$ -- similar hierarchy emerges also in \cref{ighi, ighic}. As
a consequence, the precise value of $\mmgr$ which influences
$\Ghsn$ is not so crucial for the compatibility with \Eref{bdata}
-- contrary to what happens in \cref{phi}.

\item[{\sf\small (ii)}] $\dph$ exclusively decays into the
lightest of $\rhni$'s, $\rhna$ since the upped bound of \Eref{kin}
is violated for the other $\rhni$'s.

\item[{\sf\small (iii)}] Due to the large values of $\Gysn$ and
$\GNsn$,  $\Trh$ turns out to be above $10~\PeV$ and so, the
standard scenario of the long-lived $\Gr$ and that of short-lived
$\Gr$ do not work. However, our results are perfectly acceptable
for a very short-lived $\Gr$ with $\mgr\geq40~\PeV$ and
$\mx\sim1~\TeV$ as mentioned below \Eref{m32f}.

\item[{\sf\small (iv)}] The maximal value of $|\ve_1|$ turns out
to be larger than its required value by \Eref{bdata} and for this
reason $\mrh[1]$ has to be confined close to its lower bound in
\Eref{kin} in cases A, B, C, D and G or close to the upper bound
of \Eref{kin} in cases E and F in order to decrease $|\ve_1|$ to
an acceptable level. This adjustment is accommodated by selecting
appropriately $\mD[i]$'s and the Majorana phases for any chosen
$\mn[1]$. Possible incorporation of wash-out effects
\cite{senoguz} could offer a larger flexibility in the exploration
of the available parameter space.

\end{itemize}

Let us comment, finally, on the variation of the fixed values of
parameters in our sample cases in \Tref{tab3}. Varying $\dn$ is
expected to cause a minor impact on our results due to its tiny
allowed magnitude -- see \Fref{fig3} -- and the independence of
$\msn$ from it. On the other hand, changing $\mgut$ we expect that
the various neutrino parameters in \Tref{tab3} have to be
carefully readjusted in order to reproduce the correct $\Yb$. No
dramatic changes of the $\Trh$ and $\mrh[i]$ values are expected.

The large $\Trh$ values needed in our scheme is undoubtedly a
somehow troublesome feature of our post-inflationary scenario.
This effect can be partially avoided, if we consider lower
$\mrh[i]$ values. In a such case, the explanation of the neutrino
masses via the type I seesaw mechanism will remain valid but the
leptogenesis will be inadequate. Other baryogenesis mechanisms,
such as Affleck-Dine \cite{ad} or soft \cite{soft} leptogenesis
can be alternatively activated then.



\section{Conclusions}\label{con}

We retrofitted our proposal in \cref{ighi, ighic} to obtain ample
consistency with the recent ACT DR6 extending one of the models
constructed in \cref{actpal}. Namely, our model is inspired by the
Palatini gravity and admits a realization within standard SUGRA
adopting the super- and \Kaa potentials $W$ and $K$ in
\eqs{whi}{ktot} which lead to the inflationary potential in
\Eref{Vhi} with canonical kinetic term -- cf.~\cref{un5}. The free
parameters of this potential $(\ld,M,\ck,N)$ are constrained at
the vacuum of the theory imposing the IG and gauge
coupling-unification conditions in \eqs{igc2}{mgut} besides the
inflationary constraints in \Eref{prob}. The aforementioned
requirements influenced via $\mgut$ in \Eref{mgut} our
inflarionary scenario and via the strong decay width in
\Eref{dec3} our post-inflationary scenario.

The model naturally predicts $(\ns, r)$ values within the
$1-\sigma$ margin of ACT DR6 as shown in \sFref{fig2}{b} without
any tuning and constraining the allowed margin of $\mgut$ to
values around the ``sweet'' spot predicted by the gauge
unification within MSSM -- see \Eref{res0}. As a consequence, our
present framework drastically reduces the allowed parameter space
found for a gauge-singlet inflaton with quartic potential in
\cref{actpal}. We also specified a post-inflationary completion,
based on the extra $W$ terms in \Eref{dW} and $K_2$ in \Eref{dK},
which favors split SUSY with $\mgr\simeq(40-60)~\PeV$ and allows
for baryogenesis via nTL with $\Trh$ of the order of $10~\ZeV$.
The problems with the $\Gr$ cosmology are surpassed by arranging
$\Gr$ to decay before the freeze-out of the neutralino LSP with
mass $\mx\leq2~\TeV$. In the context of split SUSY the relevant
$\mgr$ range (which is considered identical with the SUSY mass
scale $\mss$) is consistent with the Higgs-boson mass discovered
in LHC.

\acknowledgments{I would like to thank I. Antoniadis for his kind
invitation to submit my work in journal ``Astronomy'', K.
Dimopoulos, R. Kallosh and  Chia-Min Lin  for interesting
discussions.}


\appendix{Oscillating Inflation and Reheating}\label{app}

\paragraph{\hspace*{.25cm}} Trying to determine the mean
equation-of-state parameter $w$ \cite{turner, dim} after the end
of IPI we find (surprisingly) that the derived value not only
deviates hardly from that obtained within the models of chaotic
inflation but also it is close to $-1$ hinting at an additional
inflationary period. Similar findings have been already reported
for other models of inflation -- see, e.g., \cref{rhlin, wands}.
To clarify this peculiarity we solve numerically the equation of
motion for the inflaton field as detailed in \Sref{app1}, we
revaluate $w$ as described in \Sref{app2} and justify the used
formula of $\Ns$ in \Eref{prob} -- see \Sref{app3}.

\subsection{Equation of Motion} \label{app1}

To check if the period of OI \cite{os1,os2} has a sizable impact
on the post-inflationary era, we solve numerically the relevant
equation of motion for the inflaton $\sg$. In particular, the
cosmological evolution of $\sg$ in the EF is governed by the
following equation -- see, e.g., \cref{dim}:
\beq\ddot{\sg}+3\widehat{H}\dot {\sg}+\widehat V_{{\rm
IPI},\sg}=0\>\>\> \label{eqf}\eeq
where dot denotes derivation w.r.t. the cosmic time $t$, $\He$ is
the Hubble parameter in the EF and $\Vhi$ is given in \Eref{Vhi}.
In \Eref{eqf}, we neglect the damping term $\Gsn\dot{\sg}$ which
is important only at the final stage of rapid oscillations of
$\sg$ near the minimum of $\Vhi$ \cite{turner}.

The solution of Eq.~(\ref{eqf}) can be facilitated, if we use as
independent variable the (total) number of e-foldings $\vtau$
defined by
\beq \vtau=\ln \widehat R/\widehat R_{\rm i}~\Rightarrow~\dot
\vtau =\He\>\>\mbox{and}\>\>\dot \He=\He'\He\,. \label{Ndfn} \eeq
Here the prime denotes derivation w.r.t. $\vtau, \widehat R(t)$ is
the EF scale factor and $\what R_{\rm i}$ is its value at the
commencement of \fhi, which turns out to be numerically
irrelevant. Converting the time derivatives to derivatives w.r.t.
$\vtau$, Eq.~(\ref{eqf}) is equivalent to the following system of
two first order equations
\beq F_\sg =\He \what R^3 \sg'\>\>\mbox{and}\>\> \He F_\sg'=
-\Ve_{{\rm IPI},\sg}\what R^3\>\>\>\mbox{with}\>\>\>F_\sg
=\dot{\sg}\what R^3. \label{eom}\eeq
This system can be solved numerically by taking
\beq \label{Hosc} \He=\sqrt{\rho_\sg\over3\mP^2}~~\mbox{with}~~
\rho_\sg=\frac12\dot\sg^2+\what V_{\rm IPI}=\frac{F_\sg^2}{2\what
R^6}+\what V_{\rm IPI}\eeq
the energy density of $\sg$ during and after \fhi. To resolve
\Eref{eom} we impose the initial conditions (at $\vtau=0$)
$\sg(0)=(0.5-1)\mP$ and $\sg'(0)=0$. We checked that our results
are pretty stable against variation of $\sg(0)$.

To gain a deeper understanding of the $\sg$ evolution during and
after \fhi, we plot in \sFref{fig4}{a} the evolution of $\sg$ as a
function of $\vtau$ for the inputs used in \Fref{fig1}, i.e.,
$\mgut=20~\YeV$, $\dn=0$ (solid line) and $\dn=-10^{-4}$ (dashed
line). In both cases, we use $\sg(0)=0.9\mP$ and indicate the
points corresponding to $\sg(\vtau_\ast)=\sgx$ and $\sg(\vtau_{\rm
f})=\sgf$ -- the relevant $\sg$ values are listed in the Table of
\Fref{fig1}. Obviously we obtain $\Ns=\vtau_{\rm
f}-\vtau_\ast\simeq52$ in both cases. We remark that $\sg$
decreases very slowly during the IPI and then sharply from its
value $\sg_{\rm f}$ until to relax into its v.e.v in \Eref{vevs}.
Note that the last period is completed in a very limited $\vtau$
domain compared to the whole $\vtau$ region, along which \fhi\
takes place. To investigate further this last period we focus on
it in \sFref{fig4}{b} for $\dn=0$. We see there that $\sg$ passes
from the minimum of $\Vhi$ at $\vev{\sg}=0.01174$ -- see
\Fref{fig1} -- and then climbs up the hill of $\Vhi$ for a little
and oscillates backwards. This path is followed a lot of times
until $\sg$ falls finally into its minimum. The separation of the
period of OI from that of the damped oscilltion can be done by
estimating $w$ and analyzing the behavior of the decreasing
maximal $\sg$ values $\sgm$, as we do in the next section.

\subsection{Equation-of-state Parameters} \label{app2}
\begin{figure}[!t]\vspace*{-.12in}
\hspace*{-.17in}
\begin{minipage}{8in}
\epsfig{file=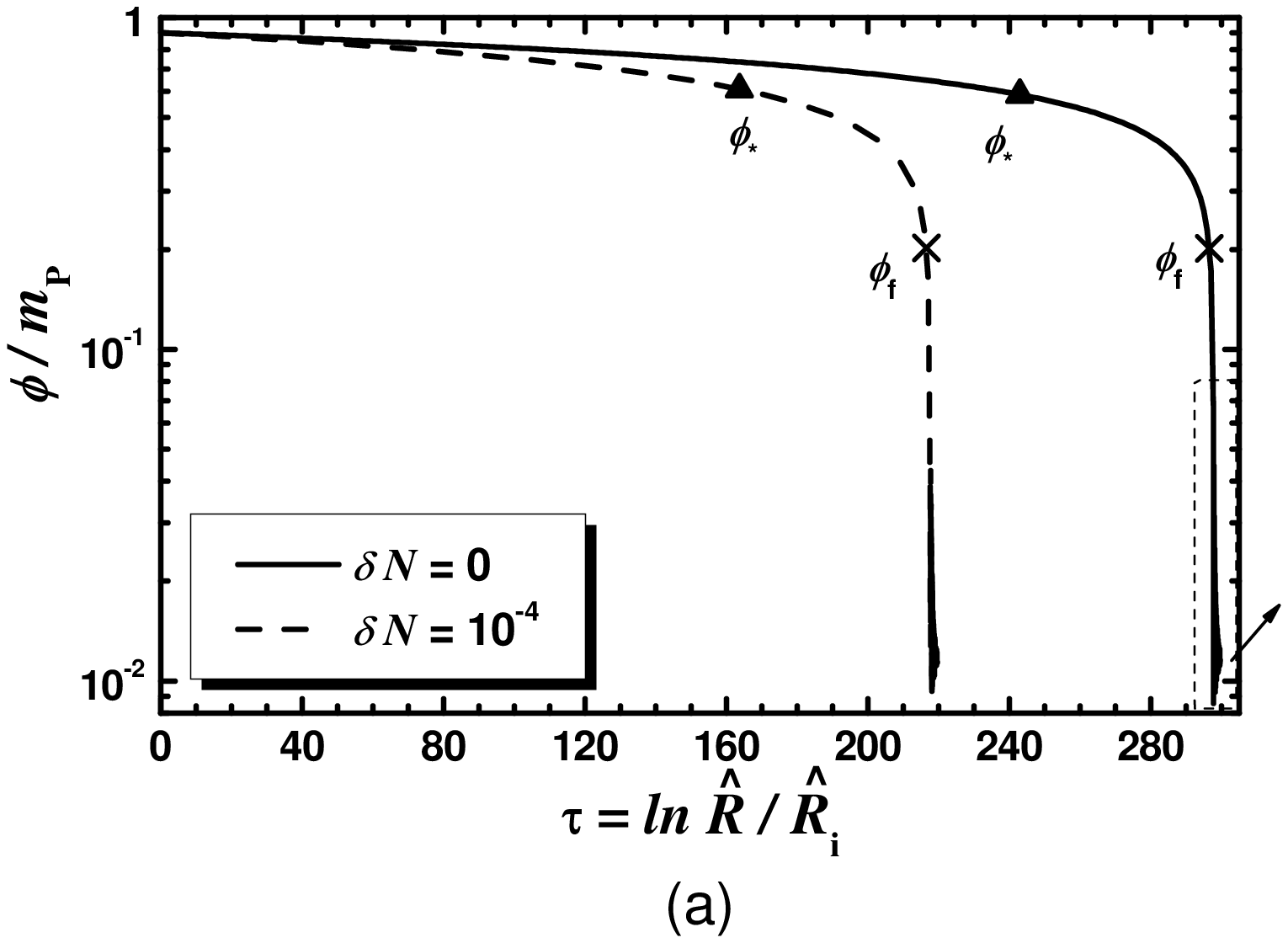,height=3.5in,angle=-90}
\hspace*{-1.cm}
\epsfig{file=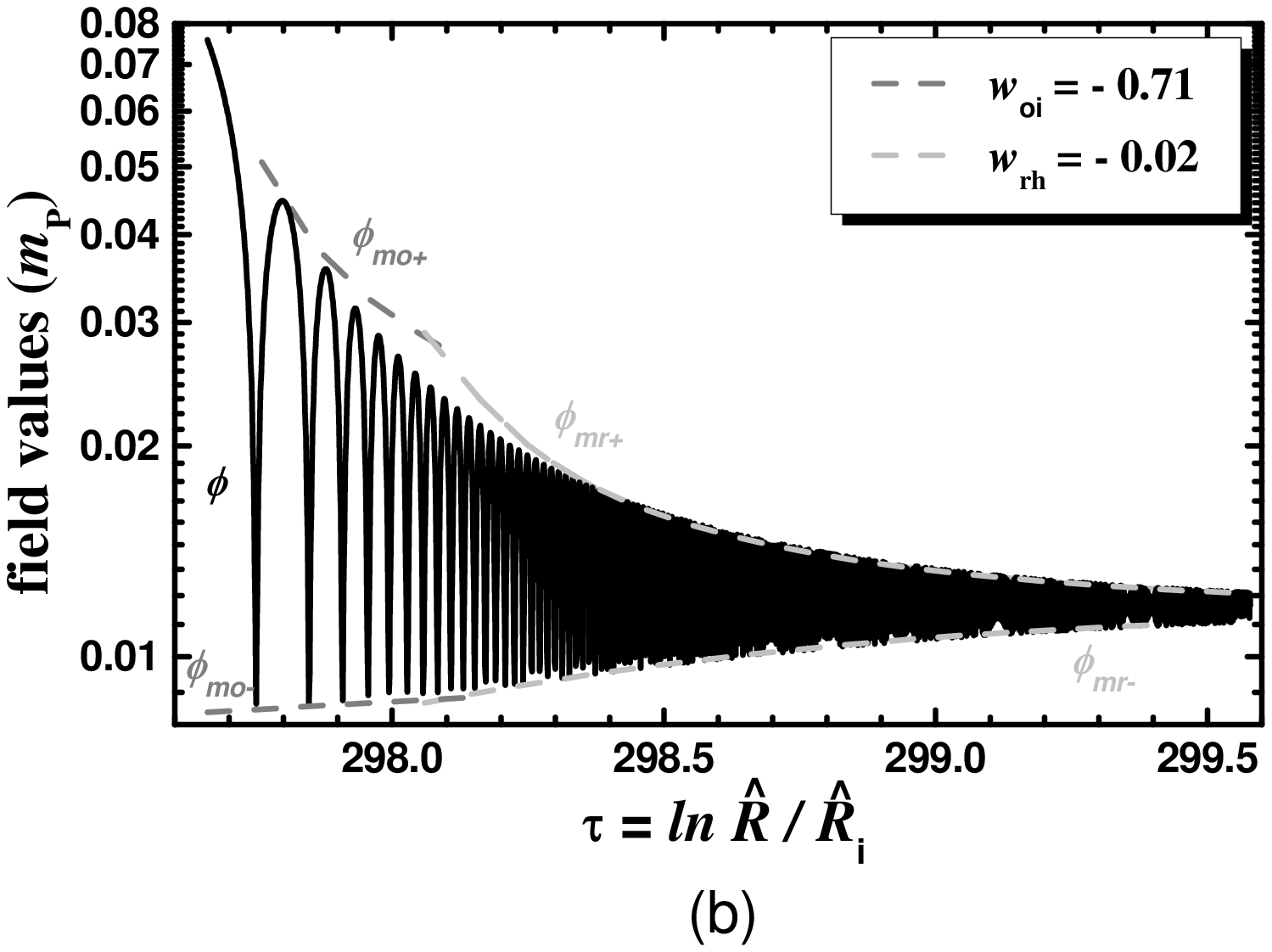,height=3.5in,angle=-90} \hfill
\end{minipage}
\hfill \caption{\sl\small Evolution of the infaton field $\sg$ as
a function of the number of e-foldings $\vtau$ during \fhi\ for
$\dn=0$ [$\dn=10^{-4}$] (solid [dashed] lines) {\sffamily\ftn (a)}
and after it for $\dn=0$ {\sffamily\ftn (b)}. The remaining inputs
are listed in the Table of \Fref{fig1}. The decreasing amplitude
of the $\sg$ oscillations for $\woi=-0.71$ [$\wrh=-0.02$] are also
depicted by gray [light gray] dashed lines {\sffamily\ftn (b)}.
}\label{fig4}
\end{figure}
We determine $w$ applying the general formula \cite{dim, review},
i.e.
\beq w=2\frac{\int_{M}^{\sgm} d\sg\,(1-
\Vhi/\Vm)^{1/2}}{\int_{M}^{\sgm} d\sg\, (1- \Vhi/\Vm)^{-1/2}}-1
\>\>\mbox{where}\>\>\Vm=\Vhi(\sgm).\label{wdef}\eeq
In general $w$ depends on the amplitude of oscillations $\sgm$
\cite{turner} and it is not just a constant as in the well-known
models of monomial potentials. The solution of \Eref{eom} gives us
the opportunity to evaluate $w$ for various $\sgm$'s and asses
which periods can be successfully described by the resulting $w$.
Indeed, according the definition above, $\rho_\sg$ in \Eref{Hosc}
has to be well approximated by \cite{turner,review}
\beq \rho_\sg(\vtau)=V_{\rm m}e^{-3(1+w)(\vtau-\vtau_{\rm
m})}\label{rhom}\eeq
for $\vtau>\vtau_{\rm m}$ where $\sg(\vtau_{\rm m})=\sgm$.
Estimating repetitively $w$ we find two values, $\woi=-0.71$ and
$\wrh=-0.02$, with $\sgm=\sg_{\rm mo}$ and $\sg_{\rm mr}$
corresponding to $\vtau_{\rm mo}=297.798$ and $\vtau_{\rm
mr}=298.6$ which reliably describe the whole $\rho_\phi$ evolution
for $\vtau\gg\vtau_{\rm f}$. Namely, we can define two branches of
$\rho_\sg$
\beqs\bea \label{rhosoi} &&\rho_{\rm oi}=V_{\rm
mo}e^{-3(1+\woi)(\vtau-\vtau_{\rm mo})}~~\mbox{for} ~~\vtau_{\rm
mo}\leq\vtau\leq\vtau_{\rm mr};\\ &&\rho_{\rm rh}=V_{\rm
mr}e^{-3(1+\wrh)(\vtau-\vtau_{\rm mr})}~~\mbox{for}
~~\vtau>\vtau_{\rm mr}.\label{rhosr}\eea\eeqs
Here $V_{\rm mo}$ and $V_{\rm mr}$ are the potential energy
density associated with the (maximal) amplitude at the onset of OI
$\vtau_{\rm mo}$ and of reheating process $\vtau_{\rm mr}$
respectively. Since for $\dot\sg=0$, $\rho_\phi\simeq\Vm$, from
the $\rho$'s in \eqs{rhosoi}{rhosr} we can find the evolution of
$\sgm$ and compare it with the solution of \Eref{eom}. Namely,
solving \Eref{Vhi} w.r.t $\sg$ we find two values for $\sgm$
\beq \sg_{\rm m\pm}=M\lf\frac{\ld}{\ld\pm4\ck\Vm^{1/2}}\rg^{1/2}.
\label{sgmpm}\eeq
Applying the previous relation for the two $\rho$'s in
\eqs{rhosoi}{rhosr} we can find the various branches of the
evolution of $\sgm$ shown in \sFref{fig4}{b} represented by
\beq \sg_{\rm mo\pm}=\sg_{\rm m\pm}(\Vm=\rho_{\rm oi})~~\mbox{for}
~~\vtau_{\rm mo}\leq\vtau\leq\vtau_{\rm mr}~~\mbox{and}~~ \sg_{\rm
mr\pm}=\sg_{\rm m\pm}(\Vm=\rho_{\rm rh})~~\mbox{for}
~~\vtau>\vtau_{\rm mr}.\eeq
We use dashed gray lines for $\sg_{\rm mo\pm}$ and light gray
dashed lines for $\sg_{\rm mr\pm}$. We remark that the first ones
provide a good covering of the $\sg$ evolution during the early
period of oscillations, whereas the second one is the envelop of
late $\sg$ oscillations. Consequently, we can conclude that $\woi$
and $\wrh$ describe reliably the periods of OI and reheating
respectively.

\subsection{Number of e-Foldings}\label{app3}

Although the duration of OI is extremely limited as shown in
\sFref{fig4}{b} it would be instructive to find its possible
contribution into the number of e-foldings $\Ns$ between horizon
crossing of the observationally relevant mode $\ks$ and the end of
IPI. We can find it as follows \cite{review}:
\bea \nonumber \frac{\ks}{H_0R_0}=\frac{H_{\star}
R_{\star}}{H_0R_0} &=&\frac{H_\star}{H_0}\frac{R_\star}{R_{\rm
If}}\frac{R_{\rm If}}{R_{\rm of}} \frac{R_{\rm of}}{R_{\rm
rh}}\frac{R_{\rm rh}}{R_{\rm eq}}\frac{R_{\rm eq}}{R_0}\\\nonumber
&=& \sqrt{V_{\rm IPI\star}\over{\rho_{\rm
c0}}}e^{-N_\star}\left({V_{\rm IPIf}\over V_{\rm
of}}\right)^\frac{-1}{3(1+\woi)}\left({V_{\rm of}\over\rho_{\rm
rh}}\right)^\frac{-1}{3(1+\wrh)} \left({\rho_{\rm
rh}\over\rho_{\rm eq}}\right)^{-1/4}\left({\rho_{\rm
eq}\over\rho_{\rm M0}}\right)^{-1/3}.\label{hor3} \eea
Here, $R$ is the scale factor in the JF, $H=\dot R/R$ is the JF
Hubble rate, $\rho$ is the energy density and the subscripts $0$,
$\star$, If, of, rh, eq and M denote values at the present, at the
horizon crossing ($\ks=R_\star H_\star$) of the mode $\ks$, at the
end of \fhi, at the end of OI, at the end of reheating, at the
radiation-matter equidensity point and at the matter domination.
More explicitly, we set $V_{\rm IPIf}=V_{\rm IPI}(\sgf)$ and
$V_{\rm of}=\rho_{\rm oi}(\vtau_{\rm mr})$. In practice, $\sg_{\rm
mr}$ can be found by comparing $\Vhi$ in \Eref{Vhi} with its
Taylor expansion about $\sg=\vev{\sg}$ in \Eref{vevs}. The largest
$\sg$ where both potential energy densities coincide gives
$\sgmr$. In our calculation we make use of the fact that $R\propto
\rho^{-1/3(1+\woi)}$ for OI, $R\propto \rho^{-1/3(1+\wrh)}$ during
reheating, $R\propto \rho^{-1/4}$ for radiation domination and
$R\propto \rho^{-1/3}$ for matter domination. For our numerical
estimates regarding the present-day quantities of the critical
energy density $\rho_{\rm c0}$ and matter $\rho_{\rm M0}$ we use
the values from \cref{act}. If we then perform the transition from
the JF to EF, taking in account that $\what R=\fr^{1/2} R$ and
$\what V_{\rm IPI}=V_{\rm IPI}/\fr^2$, we end up with the formula
for
\beq \Ns=N_\star+\frac12\ln\frac{\fr(\sgf)}{\fr(\sgx)},
\label{Nse}\eeq
shown in \Eref{prob}. Note that we ignore the discrimination
between JF and EF for $\vtau>\vtau_{\rm f}$ since at the SUSY
vacuum both systems become undistinguishable. Moreover, in the
limit where $V_{\rm of}$ approaches $V_{\rm IPIf}$ then $\Ns$
approaches the well-known formula employed \cite{review} in the
absence of OI.

In conclusion, we verified that the contribution into $\Ns$ from
the period of OI is negligible.





\def\ijmp#1#2#3{{\sl Int. Jour. Mod. Phys.}
{\bf #1},~#3~(#2)}
\def\plb#1#2#3{{\sl Phys. Lett. B }{\bf #1}, #3 (#2)}
\def\prl#1#2#3{{\sl Phys. Rev. Lett.}
{\bf #1},~#3~(#2)}
\def\rmp#1#2#3{{Rev. Mod. Phys.}
{\bf #1},~#3~(#2)}
\def\prep#1#2#3{{\sl Phys. Rep. }{\bf #1}, #3 (#2)}
\def\prd#1#2#3{{\sl Phys. Rev. D }{\bf #1}, #3 (#2)}
\def\npb#1#2#3{{\sl Nucl. Phys. }{\bf B#1}, #3 (#2)}
\def\npps#1#2#3{{Nucl. Phys. B (Proc. Sup.)}
{\bf #1},~#3~(#2)}
\def\mpl#1#2#3{{Mod. Phys. Lett.}
{\bf #1},~#3~(#2)}
\def\jetp#1#2#3{{JETP Lett. }{\bf #1}, #3 (#2)}
\def\app#1#2#3{{Acta Phys. Polon.}
{\bf #1},~#3~(#2)}
\def\ptp#1#2#3{{Prog. Theor. Phys.}
{\bf #1},~#3~(#2)}
\def\n#1#2#3{{Nature }{\bf #1},~#3~(#2)}
\def\apj#1#2#3{{Astrophys. J.}
{\bf #1},~#3~(#2)}
\def\grg#1#2#3{{Gen. Rel. Grav.}
{\bf #1},~#3~(#2)}
\def\s#1#2#3{{Science }{\bf #1},~#3~(#2)}
\def\ibid#1#2#3{{\it ibid. }{\bf #1},~#3~(#2)}
\def\cpc#1#2#3{{Comput. Phys. Commun.}
{\bf #1},~#3~(#2)}
\def\astp#1#2#3{{Astropart. Phys.}
{\bf #1},~#3~(#2)}
\def\epjc#1#2#3{{Eur. Phys. J. C}
{\bf #1},~#3~(#2)}
\def\jhep#1#2#3{{\sl J. High Energy Phys.}
{\bf #1}, #3 (#2)}
\newcommand\jcapn[4]{{\sl J.\ Cosmol.\ Astropart.\ Phys.\ }{\bf #1}, #3, no.~#4 (#2)}
\newcommand\prdn[4]{{\sl Phys. Rev. D\ }{\bf #1}, #3, no.~#4 (#2)}

\end{document}